\documentclass[twocolumn,english,reprint, superscriptaddress, breaklinks=true, showkeys, showpacs=false, nofootinbib]{revtex4}

\usepackage[T1]{fontenc}
\usepackage{color}
\usepackage{babel}
\usepackage{float}
\usepackage{amssymb}
\usepackage{listings}
\usepackage{graphicx}
\usepackage{amsmath}
\usepackage{svg}
\usepackage{booktabs} 
\usepackage[normalem]{ulem}

\newtheorem{definition}{Definition}
\newtheorem{lema}{Lemma}
\newtheorem{Theorem}{Theorem}
\newtheorem{Corollary}{Corollary}

\usepackage[unicode=true,pdfusetitle, bookmarks=true,bookmarksnumbered=false,bookmarksopen=false, breaklinks=true,pdfborder={0 0 0},backref=false,colorlinks=true]{hyperref}
\usepackage{breakurl}
\definecolor{Code}{rgb}{0,0,0}
\definecolor{Decorators}{rgb}{0.5,0.5,0.5}
\definecolor{Numbers}{rgb}{0.5,0,0}
\definecolor{MatchingBrackets}{rgb}{0.25,0.5,0.5}
\definecolor{Keywords}{rgb}{0,0,1}
\definecolor{self}{rgb}{0,0,0}
\definecolor{Strings}{rgb}{0,0.63,0}
\definecolor{Comments}{rgb}{0,0.63,1}
\definecolor{Backquotes}{rgb}{0,0,0}
\definecolor{Classname}{rgb}{0,0,0}
\definecolor{FunctionName}{rgb}{0,0,0}
\definecolor{Operators}{rgb}{0,0,0}
\definecolor{Background}{rgb}{0.98,0.98,0.98}
\lstdefinelanguage{Python}{
numbers=left,
numberstyle=\footnotesize,
numbersep=1em,
xleftmargin=1em,
framextopmargin=2em,
framexbottommargin=2em,
showspaces=false,
showtabs=false,
showstringspaces=false,
frame=l,
tabsize=4,
basicstyle=\ttfamily\small\setstretch{1},
backgroundcolor=\color{Background},
commentstyle=\color{Comments}\slshape,
stringstyle=\color{Strings},
morecomment=[s][\color{Strings}]{"""}{"""},
morecomment=[s][\color{Strings}]{'''}{'''},
morekeywords={import,from,class,def,for,while,if,is,in,elif,else,not,and,or,print,break,continue,return,True,False,None,access,as,,del,except,exec,finally,global,import,lambda,pass,print,raise,try,assert},
keywordstyle={\color{Keywords}\bfseries},
morekeywords={[2]@invariant,pylab,numpy,np,scipy},
keywordstyle={[2]\color{Decorators}\slshape},
emph={self},
emphstyle={\color{self}\slshape},
}

\makeatletter
\@ifundefined{textcolor}{}
{%
 \definecolor{BLACK}{gray}{0}
 \definecolor{WHITE}{gray}{1}
 \definecolor{RED}{rgb}{1,0,0}
 \definecolor{GREEN}{rgb}{0,1,0}
 \definecolor{BLUE}{rgb}{0,0,1}
 \definecolor{CYAN}{cmyk}{1,0,0,0}
 \definecolor{MAGENTA}{cmyk}{0,1,0,0}
 \definecolor{YELLOW}{cmyk}{0,0,1,0}
}

\hypersetup{colorlinks=true,citecolor=blue,linkcolor=cyan,urlcolor=blue,filecolor= green, breaklinks=true}
\usepackage{url}
\usepackage{breakurl}

\makeatother

\begin{document}

\author{Lucas Friedrich\href{https://orcid.org/0000-0002-3488-8808}{\includegraphics[scale=0.05]{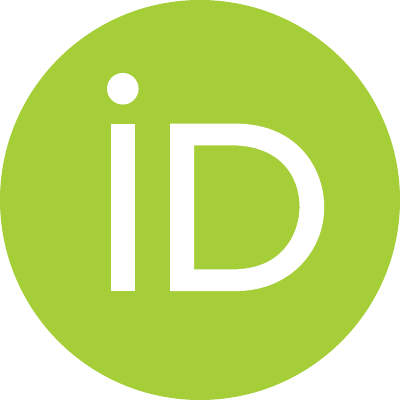}}}
\email{lucas.friedrich@acad.ufsm.br}
\affiliation{Physics Department, 
Federal University of Santa Maria, 97105-900,
Santa Maria, RS, Brazil}

\author{Tiago de Souza Farias\href{https://orcid.org/0000-0002-6697-9333}{\includegraphics[scale=0.05]{orcidid.pdf}}}
\email{tiago.farias@ufscar.br}
\affiliation{Physics Department, Federal University of S\~ao Carlos, 97105-900, S\~ao Carlos, SP, Brazil}

\author{Jonas Maziero\href{https://orcid.org/0000-0002-2872-986X}{\includegraphics[scale=0.05]{orcidid.pdf}}}
\email{jonas.maziero@ufsm.br}
\affiliation{Physics Department, 
Federal University of Santa Maria, 97105-900,
Santa Maria, RS, Brazil}

\title{Barren plateaus
are amplified by the dimension of qudits}


\begin{abstract}
Variational Quantum Algorithms (VQAs) have emerged as pivotal strategies for attaining quantum advantage in diverse scientific and technological domains, notably within Quantum Neural Networks. However, despite their potential, VQAs encounter significant obstacles, chief among them being the vanishing gradient problem, commonly referred to as barren plateaus. In this article, through meticulous analysis, we demonstrate that existing literature implicitly suggests the intrinsic influence of qudit dimensionality on barren plateaus. To instantiate these findings, we present numerical results that exemplify the impact of qudit dimensionality on barren plateaus. Therefore, despite the proposition of various error mitigation techniques, our results call for further scrutiny about their efficacy in the context of VQAs with qudits.
\end{abstract}


\maketitle
\section{Introduction}

While quantum computing has roots in the past \cite{Nielsen2010}, its substantial expansion has primarily unfolded in recent years \cite{Nature2022}. Both companies and governmental entities have made substantial investments in hardware, software, and human capital to propel the advancement of these quantum devices \cite{Riedel2019, Monroe2019}.
The pursuit of quantum computers is driven by their envisioned superiority over classical counterparts. A prominent illustration of this potential is Shor's algorithm \cite{Shor}, which was meticulously crafted for prime number factorization. This algorithm bears the capability to efficiently break cryptographic keys, thus holding profound implications across various societal realms, particularly in an interconnected world where privacy stands as a cornerstone.

While Shor's algorithm has historically been a driving force behind quantum computing development, 
the contemporary landscape, characterized by Noisy Intermediate-Scale Quantum devices (NISQ) \cite{Preskill}, has ushered in Variational Quantum Algorithms (VQAs) as the forefront strategy for achieving quantum advantages \cite{Cerezo_VQA}. These algorithms have garnered attention due to their potential to surpass classical computing methodologies. VQAs have already found applications in diverse fields, including chemical reaction simulations \cite{BAUER2020}, optimization \cite{Cerezo_VQA}, and machine learning \cite{quantum_model_multilayer_perception, kernel_methods, quantum_Convolutional, hybrid_1, hybrid_2, hybrid_3, hybrid_4}. Their versatility and efficacy in tackling intricate problems underscore their significance across a spectrum of domains.

Despite promising advancements, Variational Quantum Algorithms encounter several challenges, among which barren plateaus (BPs) stand out as a significant issue. In VQAs, a classical optimizer is employed to adjust the parameters of a quantum circuit, aiming to minimize a cost function. Typically, gradient-based methods are used to optimize these parameters, leveraging the gradient of the cost function. However, the presence of barren plateaus causes the gradient to diminish as the number of qubits increases, hindering the training process of VQAs and, as a result, constraining their practical application.

Several factors have been associated with the problem of barren plateaus, including the choice of the cost function \cite{barrenPlateaus_1}, the expressibility of the quantum circuit \cite{barrenPlateaus_2}, entanglement \cite{barrenPlateaus_4, barrenPlateaus_5}, and noise \cite{barrenPlateaus_3}. Despite research efforts to address this problem, our understanding of the phenomenon is still limited. For example, some approaches propose the use of optimization methods where the gradient of the cost function is not employed to adjust the parameters of the quantum circuit \cite{Friedrich_Evolution_strategies, lles, ANAND}. However, studies have shown that even these methods are not immune to barren plateaus \cite{barrenPlateaus_6}. As a result, other studies have been conducted to propose ways to mitigate barren plateaus \cite{barrenPlateaus_7, barrenPlateaus_8, barrenPlateaus_9, barrenPlateaus_10, barrenPlateaus_11, Friedrich_ensemble_learning}, highlighting the complexity and ongoing importance of this research area.

Quantum computers function by manipulating two-level systems known as qubits \cite{Nielsen2010}, which are analogous to the bits used in classical computing. Similar to bits, qubits possess two possible states, typically denoted as 0 or 1. However, unlike bits, each qubit can exist in a superposition of these states, allowing it to simultaneously represent both the 0 and 1 states. This property, known as superposition, is one of the fundamental characteristics that enable the quantum advantage to be achieved.

Although the leading quantum computers currently in development are designed using qubits, an alternative quantum information processing strategy is to utilize qudits. Qudits are a generalization of qubits, representing systems with a greater number of possible states, usually denoted as $d$ levels. Recent studies have explored the use of qudits in quantum computing \cite{CHI2022}, including the development of quantum machine learning algorithms based on 
VQAs \cite{ROCA2023, WACH2023, VALTINOS2023}. However, this is an area of study in its early stages, and therefore our understanding of the applicability of VQAs models using qudits is still limited, with several questions yet to be explored.

In this work, we present a new perspective on the BPs problem, examining its relationship with the dimension of qudits and introducing the concept of 
barren plateaus amplified by the dimension of the qudit. Our investigation indicates that an increase in the dimensionality of qudits correlates with a heightened impact of BPs on VQAs. Consequently, while the adoption of qudits in quantum computing presents potential benefits, the dimensional characteristics of qudits can affect the trainability of VQAs.

The remainder of this article is organized as follows. We begin by presenting a brief summary of Variational Quantum Algorithms in Section \ref{sec:vqa}. Next, in Section \ref{sec:bp}, we delve into the problem of barren plateaus. Then, in Section \ref{sec:qudit}, we introduce qudits. In Section \ref{sec:qudit_bp}, we present our main theoretical results, followed by numerical results in Section \ref{sec:result}. We conclude our findings in Section \ref{sec:conclusions}. At last, Appendix \ref{app} is used to give a detailed proof of Theorem \ref{tr:1} and Appendix \ref{appB} presents some results in log-log scale.

\section{Variational quantum algorithms}
\label{sec:vqa}

Variational Quantum Algorithms (VQAs) are emerging as a promising approach in quantum computing, offering a flexible and adaptable framework to address a wide array of complex challenges, particularly within the realm of Noisy Intermediate-Scale Quantum devices (NISQ). These devices, characterized by their smaller-scale systems, confront significant hurdles related to noise and limited qubit coherence. In typical VQAs, a parameterized quantum circuit is employed to represent a family of quantum states, with its parameters iteratively adjusted by a classical optimization algorithm. The objective is to minimize a cost function associated with a specific problem, enabling the quantum circuit to adapt and approximate the optimal solution.

In general, the cost function is defined as follows:
\begin{equation}     C(\pmb{\theta}) = Tr[O U(\pmb{\theta})\rho U(\pmb{\theta})^{\dagger}],\label{eq:cost} \end{equation}
where $\rho$ represents the system initial state, $O$ is a Hermitian operator describing an observable, and $U(\pmb{\theta})$ denotes any parametrization dependent on the parameters $\pmb{\theta}$ to be optimized.

Usually, we can define $O$ locally or globally, and, as we will see in Section \ref{sec:bp}, its choice has serious implications for the trainability of the model. On the other hand, the parameterization $U(\pmb{\theta})$ can take on a multitude of forms, which defines how the quantum gates are distributed in a quantum circuit. For example, in quantum neural networks, what generally distinguishes all the proposed models \cite{quantum_model_multilayer_perception, kernel_methods, quantum_Convolutional, hybrid_1, hybrid_2, hybrid_3, hybrid_4} is precisely the way this parameterization is obtained. Some studies have already proposed to investigate how this choice can influence the model's performance \cite{Friedrich_chip}. In Refs. \cite{RAGONE2022, LAROCCA2022}, for example, it was analyzed how to construct this parameterization based on the symmetries of the training data in the context of quantum neural networks. However, in general, this parameterization is defined as:
\begin{equation}    
U(\pmb{\theta}) = \prod_{l=1}^{L}U_{l} = \prod_{l=1}^{L}U_{l}( \pmb{\theta}_{l} ) W_{l}   ,\label{eq:parametrization} \end{equation}
where $U_{l}(\pmb{\theta}_{l})$ is a parameterization obtained from applying a sequence of quantum gates depending on the parameters $\pmb{\theta}_{l}$. The operations $W_{l}$ are another parameterizations, also obtained from applying a sequence of quantum gates, but not depending on parameters $\pmb{\theta}_{l}$, and $L$ is the depth of the parameterization $U$.

The objective of classical optimization is to obtain the parameters $\pmb{\theta}^{\ast}$ such that:
\begin{equation}     \pmb{\theta}^{\ast} = \arg_{\pmb{\theta}} \min C(\pmb{\theta}), \end{equation}
meaning that $C(\pmb{\theta})$ attains its smallest possible value. Although several optimization methods have been proposed \cite{Friedrich_Evolution_strategies, lles, ANAND}, generally, this optimization is performed using gradient-based methods. For instance, the gradient descent method involves updating the parameters of the parameterization using the gradient of the cost function. This method is described by the following optimization rule:
\begin{equation}     \pmb{\theta}^{t+1} = \pmb{\theta}^{t} - \eta \nabla_{\pmb{\theta}^t} C(\pmb{\theta}^t), \end{equation}
where $\eta$ is the learning rate. As mentioned, this is an iterative method where the parameters of $U$ are optimized to minimize the cost function; thus, $t$ refers to the current 
iteration.

\section{Barren Plateaus}
\label{sec:bp}

The problem of 
vanishing gradient in parameterized quantum circuits, also known a barren plateaus, was initially introduced in Ref. \cite{barrenPlateaus_0} in the context of quantum neural networks. In this pioneering work, the authors first demonstrated that the gradient of the cost function, used to optimize the parameters of the parameterization, exponentially decreases with the number of qubits used in the quantum circuit. Although it was observed that this problem is also related to the depth $L$ of the parameterization, it was only in Ref. \cite{barrenPlateaus_1} that theoretical results showing this relationship were obtained. The barren plateaus are defined as follows.

\begin{definition}\label{def:1}
Let the cost function be defined in Eq. \eqref{eq:cost}, with $O$ being any observable, $\rho$ the initial state of $n$ qubits, and $U$ the parameterization given in Eq. \eqref{eq:parametrization}. We say that this function exhibits barren plateaus if
\begin{equation}
     Var[\langle \partial_{k}C \rangle] \leqslant G(n) \propto \mathcal{O}\bigg( \frac{1}{b^{n}} \bigg),\ b>1.
\end{equation}
\end{definition}

This definition is obtained from Chebyshev's inequality:
\begin{equation}     Pr( |\partial_{k}C -  \langle \partial_{k}C \rangle| \geqslant \delta )  \leqslant \frac{ Var[\langle \partial_{k}C \rangle]  }{\delta^{2}}, \label{eq:Chebyshev} \end{equation}
which states that the probability that the partial derivative of the cost function with respect to any parameter $\theta_{k}$ deviates from its mean $\langle \partial_{k}C \rangle$ by a value greater than or equal to $\delta$ will be bounded by $Var[\langle \partial_{k}C \rangle]$. As shown in Refs. \cite{barrenPlateaus_0,barrenPlateaus_1}, it follows that:
\begin{equation}     \langle \partial_{k}C \rangle = 0 \ \forall \hspace{2pt} k. \label{eq:med} \end{equation}
Therefore, the inequality in Eq. \eqref{eq:Chebyshev} informs us about the probability of $\partial_{k}C$ deviating from the value $0$. As a consequence, the smaller the value of the variance, the more the value of the derivative $\partial_{k}C$ will concentrate around zero. Therefore, its trainability and consequent applicability can be severely hindered.

Several results in the literature have already related barren plateaus to different properties \cite{barrenPlateaus_0, barrenPlateaus_1, barrenPlateaus_2, barrenPlateaus_3, barrenPlateaus_4, barrenPlateaus_5}, such as expressiveness, which is related to the capacity of the parameterization $U$ to access the Hilbert space. The greater its capacity to access this space, the more expressive it will be. For example, in Ref. \cite{barrenPlateaus_2} the authors analyzed the relationship between expressiveness and barren plateaus. They showed that the higher the expressiveness of the parameterization, the greater will be its affects on barren plateaus. Recently, several studies have aimed to analyze how VQAs are affected by expressiveness \cite{FRIEDRICH_Expre_2023, SIM2019, HUBREGTSEN2021}.

In Ref. \cite{barrenPlateaus_1}, the authors analyzed how barren plateaus can be influenced by the choice of the cost function. The cost function from Eq. \eqref{eq:cost} depends on $O$, which generally can be any observable. The authors investigated how the choice of this observable affects barren plateaus. They considered two different types of observables: global observables and local observables. A global observable is defined so that the value of the cost function depends on the simultaneous measurement of all qubits. On the other hand, a local observable is defined when the value of the cost function depends only on the measurement of one qubit or some pairs of qubits. As shown in Ref. \cite{barrenPlateaus_1}, the cost function will not exhibit the problem of barren plateaus in the second case, of local observables, if the relationship between the depth of the parameterization and the number of qubits is $\mathcal{O}(1)$ or $\mathcal{O}(\log(n))$. However, for the first case of global observables, the cost function will always exhibit the problem of barren plateaus.

\section{Qudits}
\label{sec:qudit}

Qudits represent a generalization of qubits, where a qubit is a particular case of a qudit with dimension $d=2$. A $d$-dimensional qudit state can be described in terms of the standard basis $\{ |0\rangle,|1 \rangle, \cdots ,| d-1 \rangle \}$:
\begin{equation} |\psi \rangle = \sum_{l=0}^{d-1} \alpha_{l}|l\rangle, \end{equation}
with $\sum_{l=0}^{d-1} |\alpha_{l}|^{2}=1$. Although neutral multilevel Rydberg atoms or molecular magnets may be considered the best candidates for implementing qudits \cite{Weggemans2022}, most advancements in quantum information have been achieved using photons \cite{Wang2023}. In these cases, the state of the qudit is represented by a single photon superposed over $d$ modes, which can be spatial, temporal, frequency, or orbital angular momentum modes.

In quantum computing based on qubits, the state $|\psi \rangle$ describing a system of $n$ qubits is manipulated through quantum operations, such as the Hadamard gate, CNOT gate, and rotation gates, among others \cite{Wang2020}. With appropriate adjustments, it is also possible to define quantum gates specifically designed to work with systems involving qudits. For instance, for the purposes of this work, two gates that can be adapted to operate with qudits are the rotation gate 
\begin{equation}
    R_{jk}^{\alpha} = e^{-i\theta S_{\alpha}^{jk}/2},
\end{equation}
where the generalized Gell-Mann matrices are \cite{GGM}
\begin{align}
    & S_x^{jk} = |j\rangle\langle k| + |k\rangle \langle j| \text{ with } 1 \leqslant j<k \leqslant d', \label{eq:Sx} \\
   &  S_y^{jk} = -i|j\rangle\langle k| + i|k\rangle \langle j| \text{ with } 1 \leqslant j<k \leqslant d',\label{eq:Sy} \\
   & S_z^{j} = \sqrt{\frac{2}{j(j+1)}}\sum_{k=1}^{j+1} (-j)^{\delta(k,j+1)}|k\rangle \langle k|, \label{eq:Sz}
\end{align}
with $1 \leqslant j \leqslant d'-1$ for $S_z^{j}$, and the CNOT gate
\begin{equation}
    CNOT |x\rangle |y\rangle = 
         |x\rangle |x+y (\bmod d)\rangle, 
\end{equation}
with $x,y=0,\cdots,d-1$. 
Here we use $d'$ instead of $d$ to denote the dimension of the qudit, since we will use $d$ to indicate the dimension of the composite system.

In recent years, an increasing number of studies have explored the potential advantages of qudit-based quantum computing \cite{2023Fischer, 2024nikolaeva}. For instance, qudits can enable more efficient quantum error correction schemes by leveraging higher-dimensional encodings, which can improve fault tolerance and resilience to noise \cite{2025keppens}. Additionally, they offer enhanced capabilities for characterizing multipartite quantum systems, facilitating a richer description of quantum correlations and entanglement structures that are inaccessible to qubit-based approaches \cite{2013huber}. 

Another advantage is the potential for faster gate decompositions, where qudit-based operations can reduce the depth of quantum circuits and streamline computational processes \cite{2019gokhale}. Furthermore, in variational quantum algorithms such as the Variational Quantum Eigensolver, qudits can reduce the required number of computational units by encoding more information per quantum register, albeit at the cost of operating in a higher-dimensional Hilbert space \cite{2024kim}.

Even when qudit-based circuits must ultimately be decomposed into qubit operations for execution on conventional qubit-based quantum hardware, qudits can still offer advantages. For instance, in the case of state preparation for classical data in Variational Quantum Algorithms, the input data is rarely binary, necessitating an appropriate encoding into quantum states \cite{2024rath, 2024ranga}. By leveraging qudits, we can achieve a more natural and intuitive representation of such data, reducing the complexity of the encoding process and potentially improving the efficiency of quantum algorithms \cite{2024Mandilara}.

Although promising, quantum computers are predominantly in the development phase, focusing on qubit operations. Furthermore, the potential use of qudits in solving VQAs problems is an emerging area, and our theoretical foundations are still considerably limited. In the next section, we aim to expand these foundations by examining how barren plateaus are influenced by the dimension of qudits.

\section{Barren plateaus in qudit systems}
\label{sec:qudit_bp}

In this section, we will discuss how BPs are 
affected by the dimension of qudits. To achieve this, we begin by presenting the following definition:
\begin{definition}\label{def:2}
    (Barren plateaus 
    amplified by the dimension of the qudits) Let the cost function be defined in Eq. \eqref{eq:cost} with $U$ given by Eq. \eqref{eq:parametrization} and $O$ any observable. We say that this function suffers from 
    dimension-amplified BPs if
    \begin{equation}
        Var[\langle \partial_{k}C \rangle] \leqslant F(n,d')\propto \mathcal{O}\bigg( \frac{1}{d'^{n}} \bigg),
\end{equation}
with $d' \geqslant 2$ being the dimension of the qudits and $n$ the number of qudits.
\end{definition}

Thus, alike to Definition \ref{def:1}, the Definition \ref{def:2} is obtained from Chebyshev's inequality and tells us that if the variance decreases as the dimension $d'$ of the qudits increases, then 
the VQAs will have BPs polynomially amplified by the dimension of the qudits.

To prove 
that barren plateaus are amplified by the dimension of the qudit, we will consider the formalism of $t-$designs. The $t-$designs are defined as follows: Consider a finite set $\{W_{y}\}_{y \in Y}$ (of size $|Y|$) of unitaries with a Hilbert space of dimension $d$. If $P_{(t,t)}(W)$ is an arbitrary polynomial of degree at most $t$ in the elements of the matrix of $W$ and at most $t$ in $W^{\dagger}$, and 
\begin{equation}     \frac{1}{| Y |} \sum_{y \in Y} P_{(t,t)}(W_{y}) = \int_{U(d)}d\mu(W)P_{(t,t)}(W),\label{eq:tdesign} \end{equation}
we say that this finite set is a $t-$design. This result implies that the average of $P_{(t,t)}(W)$ over the $t-$design is indistinguishable from integration over $U(d)$ with respect to the uniform-Haar distribution \cite{2_design}.

From this definition of $t-$design, we are able to derive several lemmas that are useful when obtaining theoretical results showing that indeed the variance of the partial derivative of the cost function in Eq. \eqref{eq:cost} decreases as the number of 
qubits increases. However, we also observe that we can use this definition of $t-$design to show that the dimension of qudits also amplifies barren plateaus. To demonstrate this, we simply observe that this definition holds for any set of unitaries of degree $d$.

The variable $d$ is commonly defined as $d = 2^{n}$, where $n$ represents the number of qubits used in the quantum circuit, and the base $2$ corresponds to the binary dimensionality inherent to qubits. When extending this framework to accommodate qudits, the primary modification involves substituting the base $2$ with the dimension $d'$ of the qudits. Consequently, $d$ is redefined as $d = d'^{n}$. This adaptation allows the application of existing lemmas that explore the relationship between variance and the number of qubits to also assess the impact of qudit dimensionality.

To analyze how the variance behaves as the dimension $d'$ of the qudits increases, we must first obtain an expression for the derivative of $C$. Accordingly, we start by rewriting the parametrization $U$ given by Eq. \eqref{eq:parametrization} as:
\begin{equation}
    U = U_{L}U_{R}, 
\end{equation}
with 
\begin{equation}
    U_{L} = \prod_{l=1}^{p-1}U_{l}W_{l}
\end{equation}
and
\begin{equation}
    U_{R} = \prod_{l=p}^{L}U_{l}W_{l},
\end{equation}
where $U_{l}$ will be given by
\begin{equation}
    U_{l} = \bigotimes_{m=1}^{n}R_{jk}^{\alpha}(\theta_{ml}).\label{eq:Ul}
\end{equation}

Thus, from Eq. \eqref{eq:cost}, we have
\begin{equation}
    \partial_{k}C = \frac{\partial C}{ \partial \theta_{qp} } = \frac{i}{2} Tr \bigg[ U_{L}^{\dagger} O U_{L} \big[ U_{R}\rho U_{R}^{\dagger}, [I_{q} \otimes S_{\alpha}^{jk} ] \big] \bigg],
\end{equation}
where $S_{\alpha}^{jk}$ is given by Eqs. \eqref{eq:Sx}, \eqref{eq:Sy}, and \eqref{eq:Sz}. For more details on obtaining this derivative, see Appendix \ref{app}.

With this derivative, we can see that if $U_{L}$ or $U_{R}$ form a $1$-design, then
\begin{equation}
    \langle \partial_{k} C \rangle = 0 \text{  } \forall k.
\end{equation}
Therefore, the smaller $Var[\partial_{k}C]$, the closer $\partial_{k}C$ will be to the value $0$, making it difficult to train the model. The same holds if $U_{L}$ and $U_{R}$ simultaneously form a $1$-design. For more details on how to obtain this result, see  Appendix \ref{app}. 
Next, we will present Theorem \ref{tr:1}, which relates the variance of $\partial_{k}C$ to the dimension $d'$ of the qudits.

\begin{figure*}
    \centering
    \includegraphics[scale=0.8]{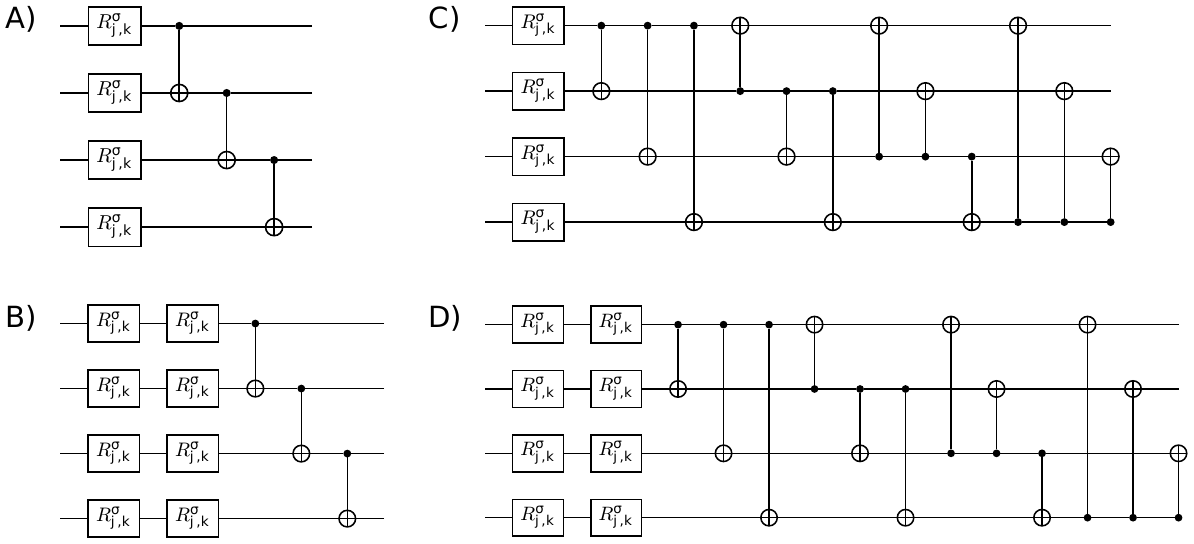}
    \caption{Illustration of the parameterizations used in this article. In this figure, each parameterization shows the form of the unitary $U_{l}$ used in Eq. \eqref{eq:parametrization}. For parameterizations A and B, the CNOT gate is applied only between neighboring pairs of qudits, while in the parameterizations C and D it is applied between all pairs of qudits. In all these parameterizations, rotation gates $R^{\sigma}_{(j,k)}$ are applied to all qudits. The pair of variables $(j,k)$ are the indices of the Gell-Mann matrices, that we randomly choose. 
    Moreover, $\sigma$ indicates which axis the rotation gate will be applied to, with $\sigma = X, Y, Z$. During the simulations, the $\sigma$ values are also chosen at random. }
    \label{fig:param}
\end{figure*}

\begin{Theorem}\label{tr:1}
Let the cost function be defined in Eq. \eqref{eq:cost}, with $O$ being any observable, $\rho = | \psi \rangle \langle \psi |$ an initial state, and $U$ the parameterization defined in Eq. \eqref{eq:parametrization}, with $U_{l}$ given by Eq. \eqref{eq:Ul}. Then the variance of the partial derivative of the cost function in Eq. \eqref{eq:cost} with respect to any parameter $\theta_{k}$ will be:
\begin{equation}
    Var[\partial_{k}C] = \frac{ d'^{(n-1)} }{d+1} \bigg( \frac{ Tr[O^{2}] }{ d^{2}-1 } - \frac{ Tr[O]^{2} }{ d(d^{2}-1) } \bigg),
\end{equation}
where $d=d'^{n}$ with $d'$ being the dimension of the qudits and $n$ the number of qudits used in the model.
\end{Theorem}

The proof of this theorem is presented in Appendix \ref{app}. Below, we present a corollary derived from this theorem, in which we analyze the behavior of the variance in a particular case. This specific case will be used for a numerical analysis to verify the validity of the theorem.

\begin{Corollary}\label{cor:1}
Let the cost function be defined in Eq. \eqref{eq:cost} with $O = |0 \rangle \langle 0|$. From Theorem \ref{tr:1}, we have:
\begin{equation}
    Var[\partial_{k}C] = \frac{1}{ d'(d'^{n} +1 )^{2} }.
\end{equation}
Therefore, from this result, we see that the variance will decrease as the dimension $d'$ of the qudits used in the VQA increases.
\end{Corollary}

\section{Results}\label{sec:result}

In this section, we present the numerical results confirming our theoretical findings, demonstrating numerically that indeed the dimension of the qudits induces the problem of barren plateaus. To perform these simulations, we used the PyTorch library \cite{Paszke} to implement a series of operations involving VQAs with qudits. For more details about these operations, please refer to the code provided at the link appearing in the Data Availability statement.

For these results, we used the general form of parameterization given by Eq. \eqref{eq:parametrization}, where $U_{l}$ is determined by the parameterizations illustrated in Fig. \ref{fig:param}. All these parameterizations depend on the CNOT gate and on the rotation gate $R^{\sigma}_{j,k}$ with $\sigma = X,Y,Z$.

\begin{figure*}
    \centering
    \includegraphics[scale=0.4]{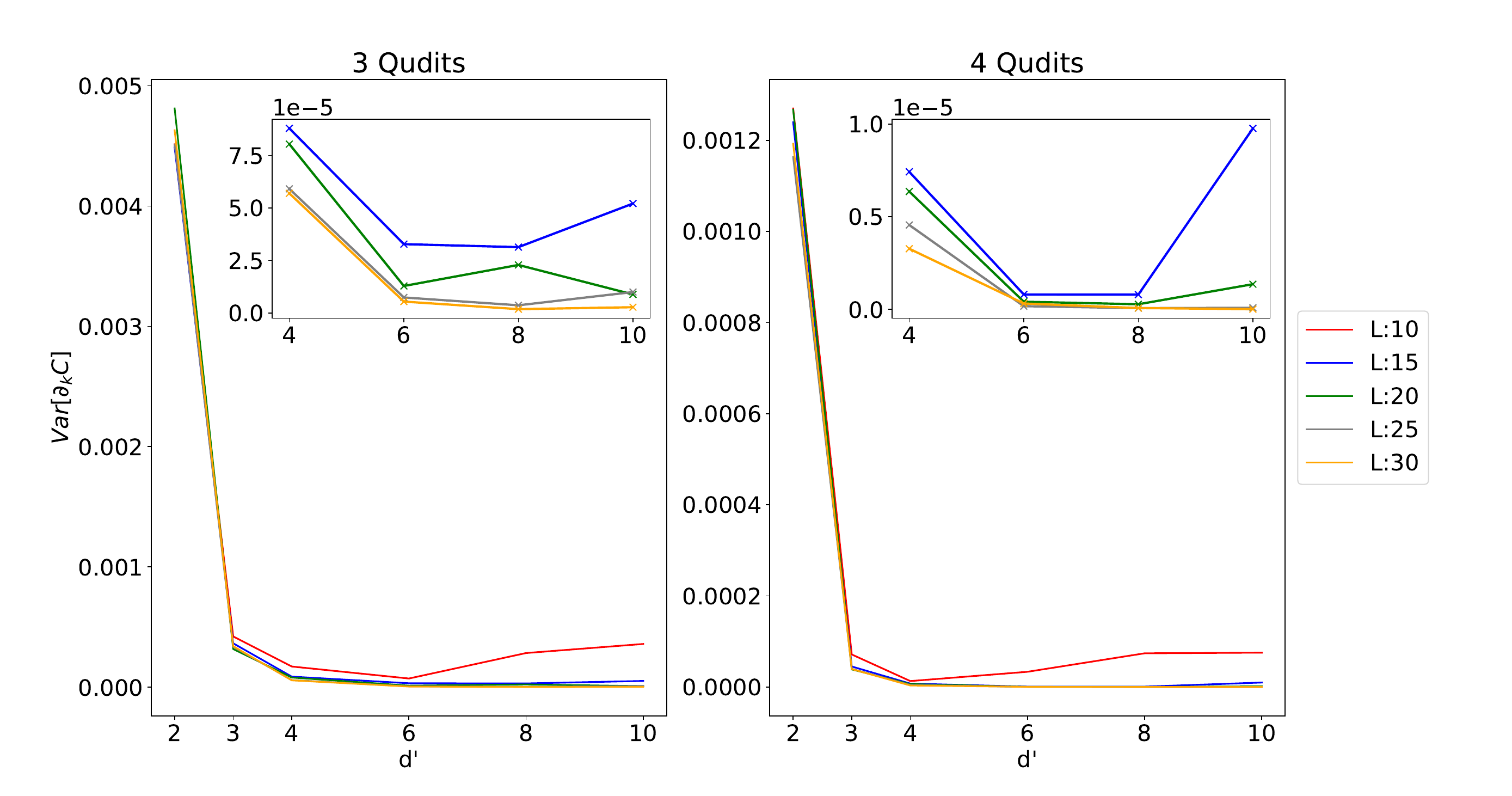}
    \caption{
    Behavior of the variance of the cost function, Eq. \eqref{eq:cost}, with $O=|0 \rangle \langle 0|$ and parameterization $U$, Eq. \eqref{eq:parametrization}, with $U_{l}$ given by Fig. \ref{fig:param}A. In this case, we can see that for sufficiently large $L$, the behavior of the variance is in accordance with Theorem \ref{tr:1}. However, for low values of $L$, specifically for $L=10$ and $L=15$, we can see that the behavior of the variance differs from what is expected according to Theorem \ref{tr:1}. As we will discuss later, this 
    happens because the parameterization set $U$ generated does not form an exact $t$-design but rather an approximation. So, it is expected that the behavior of the variance differs from the theoretical result.}
    \label{fig:model_1}
\end{figure*}

\begin{figure*}
    \centering
    \includegraphics[scale=0.4]{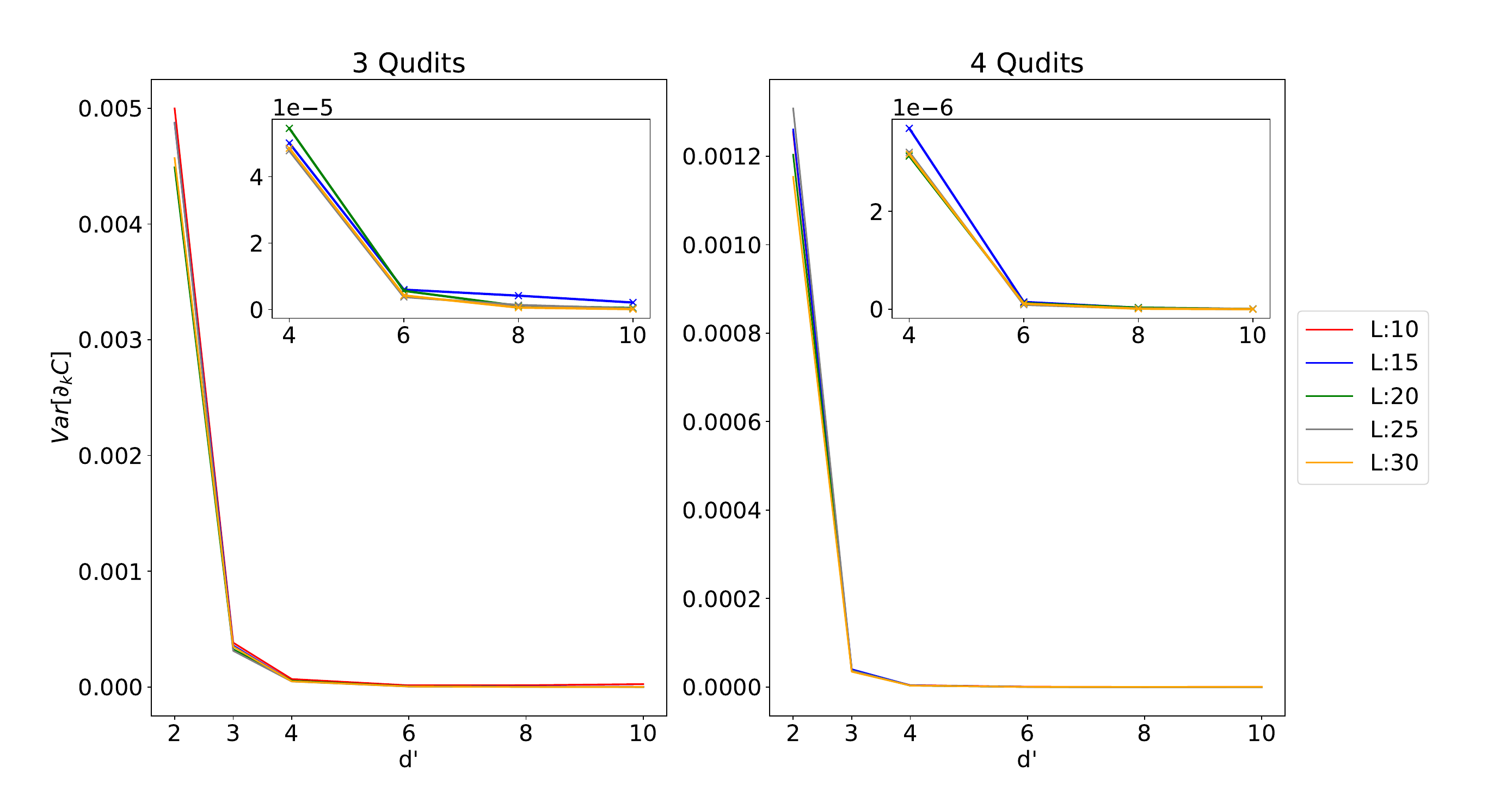}
    \caption{
    Behavior of the variance of the cost function, Eq. \eqref{eq:cost}, with $O= |0 \rangle \langle 0| $ and parameterization $U$, Eq. \eqref{eq:parametrization}, with $U_{l}$ shown in Fig. \ref{fig:param}B. In contrast to the previous case (Fig. \ref{fig:model_1}), 
    here the behavior of the variance of the cost function is in accordance with Theorem \ref{tr:1} in all cases.}
    \label{fig:model_2}
\end{figure*}

\begin{figure*}
    \centering
    \includegraphics[scale=0.4]{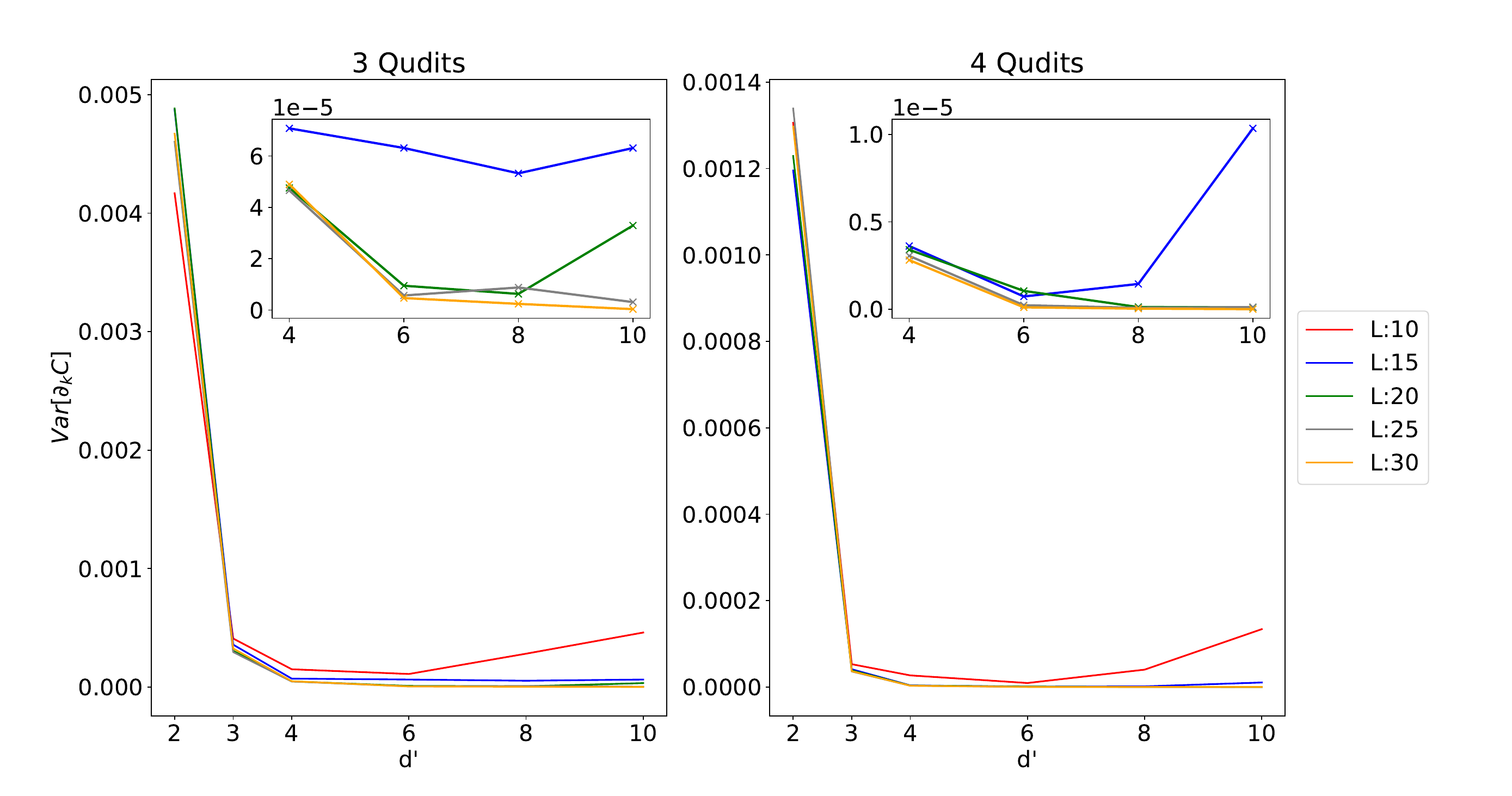}
    \caption{Behavior of the variance of the cost function for $U_{l}$ given by Fig. \ref{fig:param}C. Similar to the case seen in Fig. \ref{fig:model_1}, for relatively low values of $L$, the behavior of the variance differs from what is expected according to Theorem \ref{tr:1}.
}
    \label{fig:model_3}
\end{figure*}

\begin{figure*}
    \centering
    \includegraphics[scale=0.4]{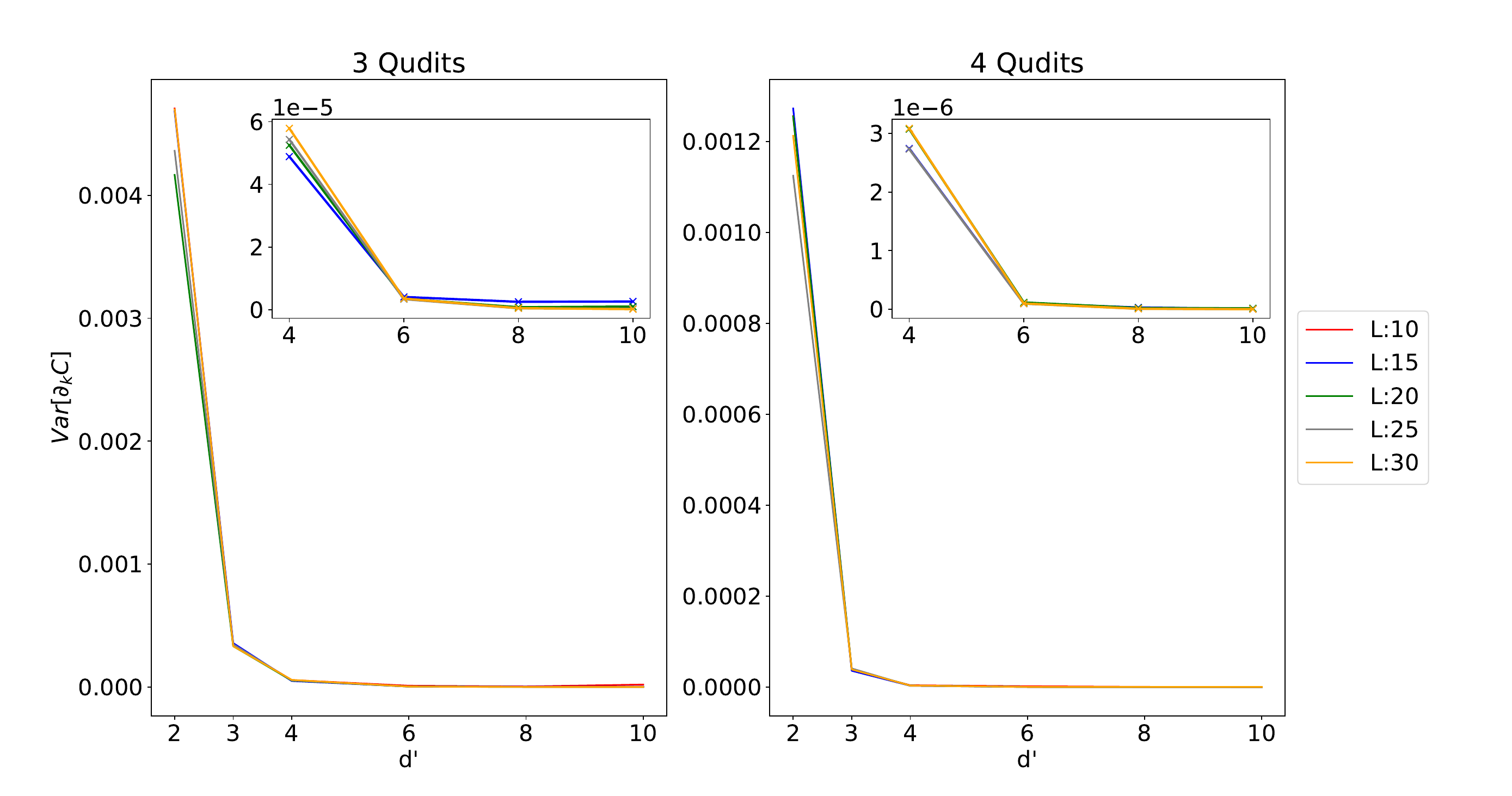}
    \caption{
    Behavior of the variance of the cost function for the parameterization obtained using $U_{l}$ shown in Fig. \ref{fig:param}D. Again, we use $O= |0 \rangle \langle 0|$. As we can observe immediately, similar to the case seen in Fig. \ref{fig:model_2}, the behavior of $Var[\partial_{k}C]$ is in accordance with Theorem \ref{tr:1} in all cases analyzed.}
    \label{fig:model_4}
\end{figure*}



The variance of $\partial_{k}C$ with respect to each $U_{l}$ illustrated in Figure \ref{fig:param} was obtained following a specific process. Initially, we selected the desired form to generate $U_{l}$. Then, we generated 2000 random parameterizations $U$ to calculate the variance of $\partial_{k}C$. Each of these parameterizations $U$ was obtained as follows: first, we defined a depth $L$ and the number of qudits to be used in the parameterization. Subsequently, we determined the dimension $d'$ of the qudits. Next, we randomly selected each rotation gate used in the parameterization, i.e., we chose $\sigma$ randomly in the set $\{X,Y,Z\}$. If $\sigma$ was chosen as $X$ or $Y$, we randomly generated the pair of variables $(j,k)$ such that $1 \leqslant j < k \leqslant d'$. In the case where $\sigma = Z$, we randomly generated $j$ such that $1 \leqslant j \leqslant d'-1$. Finally, we generated the parameters $\pmb{\theta}$ from a uniform distribution in the interval $[0,2\pi]$. With these parameters, we calculated the derivative $\partial_{k}C$ for $k=(1,1)$, i.e., the derivative with respect to the rotation gate acting on the first qudit of the first layer.

As an observable, we chose $O = |0 \rangle \langle 0|$. Thus, from Corollary \ref{cor:1}, we know that the variance in this case tends to decrease as the dimension $d'$ of the qudits increases. In the following results, we explore different values of $L$, aiming to analyze how the variance is affected by the depth of the parameterization. Additionally, we focus on analyzing the behavior of the variance for parameterizations obtained using $3$ and $4$ qudits. This limited selection is justified for two main reasons. Firstly, our central goal is to demonstrate that barren plateaus are 
amplified by the dimension of the qudits, thus our results focus on analyzing the relationship between the variance of $\partial_{k}C$ and the dimension of the qudits. Secondly, hardware limitations also play a significant role. Generally, when performing VQA simulations on classical computers, the maximum number of qubits we can simulate is already limited due to the exponential growth of computational resources required as the number of qubits increases. In the case of qudits, this challenge is even greater, especially for relatively large values of $d'$.

Figure \ref{fig:model_1} illustrates the behavior of the variance for the parametrization $U_{l}$ as depicted in Figure \ref{fig:param}A. It is evident that the variance generally decreases with an increase in the qudit dimension $d'$, particularly for $L=25$ and $L=30$. However, in scenarios where $L=10, 15, \text{ and } 20$, the variance predominantly decreases, yet there are instances where it unexpectedly increases. Two primary hypotheses are proposed to explain this phenomenon:

\begin{enumerate}
    \item The number of generated parameterizations $U$ to calculate the variance is limited. In fact, the possible number of parameterizations is unlimited, since each parameter $\theta_{k}$ is a continuous value. Therefore, even though we used 2000 parameterizations, this is still a very low number. As a result, the 
    obtained variance is 
    an under-sampled estimate.

    \item The theoretical results 
    are based on the assumption that the set of generated parameterizations $U$ forms a $t-$design. However, this set does not form an exact $t-$design but rather an approximation. Therefore, it is expected that for these cases, the variance differs to some extent from the theoretical results.
\end{enumerate}

Regarding the number of qudits, although we only used two values, it is possible to see that the variance using $4$ qudits is lower than the one obtained using $3$ qudits. This is in accordance with Theorem \ref{tr:1}, where we observe that indeed the number of qudits will influence the variance, and the larger this number of qudits is, the lower will be the variance.

Figure \ref{fig:model_2} shows the behavior of the variance for the parametrization illustrated in Figure \ref{fig:param}B. In contrast to the previous case, where an increase in variance was observed in some instances, in this case all variances decreased as the dimension $d'$ of the qudits increased. Additionally, we noticed that the variance with $4$ qudits is lower than the variance with $3$ qudits. Therefore, in this case, the behavior of the variance is in complete agreement with the expected results according to Theorem \ref{tr:1}.

Next, in Figure \ref{fig:model_3}, we analyze the behavior of the variance in the case where the parametrization $U_{l}$ is illustrated in Figure \ref{fig:param}C. Immediately, we observe that, similar to the case shown in Figure \ref{fig:model_1}, there were instances where the variance increased when it was expected to decrease. Again, the two possible explanations for this behavior are: the number of $U$ parametrizations used to calculate the variance is insufficient, which can lead to under-sampling; or the fact that the set of unitaries obtained does not form an exact $t-$design, but rather an approximation, which can result in a variance diverging from the expected theoretical result. Additionally, we observe that for sufficiently large values of $L$, the behavior of the variance is in accordance with Theorem \ref{tr:1}, and the variance with 4 qudits was lower than the variance obtained with 3 qudits, as expected.

Finally, in Figure \ref{fig:model_4}, we observe the behavior of the variance for the case where $U_{l}$ was given by Figure \ref{fig:param}D. Initially, we note that in all cases the variance decreased as the dimension $d'$ of the qudits increased, and the variance obtained when using 4 qudits is lower than that obtained when using 3 qudits. Thus, we can conclude that the variance behaved as expected by Theorem \ref{tr:1}.

As we have seen, there were cases, Figures \ref{fig:model_1} and \ref{fig:model_3}, where the behavior of the variance was different from what was expected according to Theorem \ref{tr:1}. Initially, we presented two possible explanations for this behavior. However, upon analyzing all the results obtained, we can conclude that the best explanation for this behavior is that the set of unitaries generated does not form, in these cases, an exact $t-$design but rather an approximation. Therefore, it is expected that the behavior of the variance differs to some extent from the theoretical result.

To justify this conclusion, we first ruled out the first possible explanation, which is based on statistical under-sampling due to the limited number of $U$ parametrizations used to calculate the variance. However, if this was the case, then the same behavior should be observed in Figures \ref{fig:model_2} and \ref{fig:model_4}. Additionally, we should see a much more complex behavior for the cases shown in Figures \ref{fig:model_2} and \ref{fig:model_4} if this were the correct explanation, because in these cases the total number of possible parametrizations is higher than the total number for the parametrizations obtained using Figs. \ref{fig:param}A and \ref{fig:param}C, thus the numerical error should be more evident for these cases.

With the possibility of under-sampling ruled out, another hypothesis is that the set of unitaries does not form an exact $t-$design. Therefore, it is expected that the behavior of the variance differs to some extent from the theoretical result.



\begin{figure}
    \centering
    \includegraphics[scale=0.59]{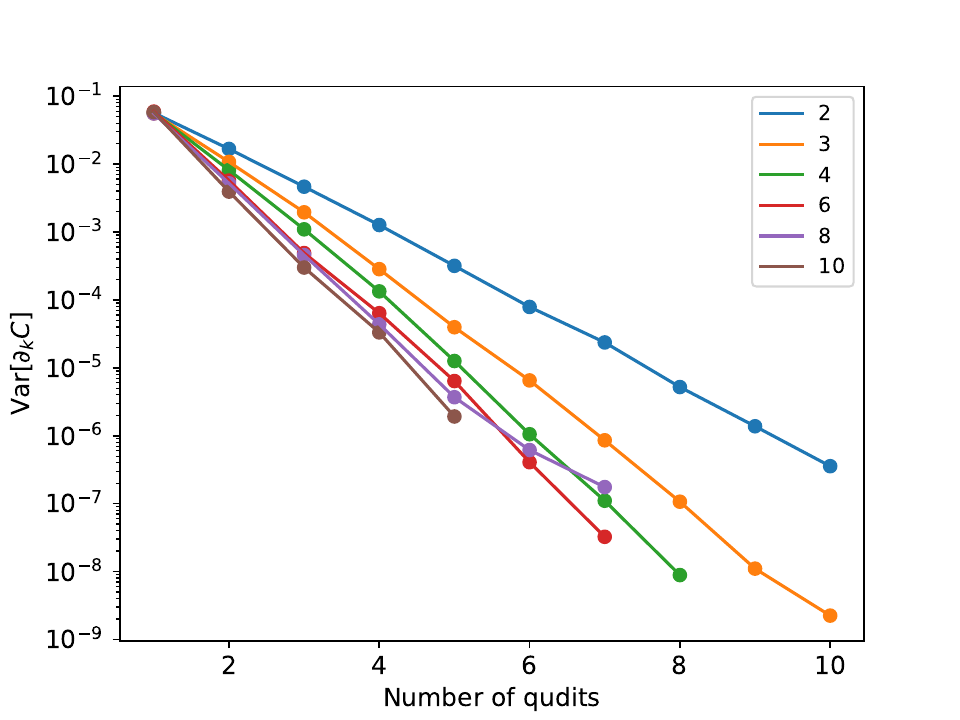}
    \caption{Exponential decay of the gradient variance as a function of the number of qudits for the parametrization in Fig. \ref{fig:param}A with 10 layers. The curves represent different qudit dimensions, illustrating that higher-dimensional qudits amplified the barren plateaus phenomenon by accelerating the decay in variance.}
    \label{fig:qudits_dim}
\end{figure}

In addition to our primary analysis, we conducted a numerical investigation into the barren plateaus phenomenon as a function of the number of qudits in the system, as well as the influence of qudit dimensionality. Figure \ref{fig:qudits_dim} shows the variance of the gradient with respect to the number of qudits for the $U_l$ presented in the figure \ref{fig:param}A,  using 10 layers and varying qudit dimensions. The results reveal an exponential decay in the gradient variance as the number of qudits increases. Furthermore, the qudit dimension significantly impacts this behavior, with higher-dimensional qudits accelerating the decay rate of the gradient variance. This observation highlights the compound challenge of training quantum circuits with increasing the system size and qudit dimensionality, as the barren plateaus effect becomes more pronounced in higher-dimensional qudit systems.

Nevertheless, since the variance of the gradient decreases polynomially with the dimension of the qudit (as reinforced by the log-log scale plots presented in Appendix \ref{appB}), as opposed to the exponential decay observed with the number of qudits, this property can be exploited when leveraging higher-dimensional qudits. This observation highlights the significance of data encoding strategies and gate decomposition in quantum circuits. If certain data representations are more naturally suited to qudit-based encodings, and an optimal number of qudits can be identified, or alternatively, if the system can be better decomposed into qubits in a manner that preserves a slower variance decay, then this approach can serve as a means to mitigate the gradient decay and, consequently, alleviate the effects of the barren plateaus phenomenon.

\section{Conclusions}\label{sec:conclusions}

Recently, some studies have proposed the use of variational quantum algorithms (VQAs) with qudits to solve machine learning problems. Although this approach is promising, it is still in its early stages of development. As a result, our understanding of issues such as the trainability of these models is still limited. In this work, our goal was to analyze how 
the barren plateaus
phenomena is affected by the qudit dimension. By using the formalism of $t-$designs, we observed BPs amplified by the dimension of the qudits. This comes from the fact that, from the formalism of $t-$designs, we observe a dependence on the dimension $d$ of the Hilbert space. This dimension is $d = 2^n$ for qubits. However, for qudits, we can generalize this dimension to $d = d'^n$, where $d'$ represents the dimension of each qudit.

Based on this analysis, we derived Theorem \ref{tr:1}, which establishes a relationship between the variance of the partial derivative of the cost function and the dimension $d'$ of the qudits. To confirm this result, we conducted a series of numerical experiments in which we analyzed how the variance of the partial derivative of the cost function behaves for different values of $d'$, using various parameterizations. The results obtained confirm that indeed, barren plateaus are 
amplified by the dimension of qudits.

Furthermore, the results obtained allow us to open a discussion on how the mitigation methods proposed so far can be applied in this specific case. Generally, these methods aim to increase the variance of the cost function. For example, consider a problem that can be solved by a VQA using $L$ layers and $n$ qubits, and suppose that, for this specific case, the variance is equal to a value $x$. Typically, mitigation methods simply increase this value to $x + \delta$, where $\delta \ll 1$. However, when using a VQA with qudits, we observe that the value of the variance is significantly lower than that obtained with a VQA with qubits. This leads us to question whether this value $\delta$ is really sufficient to mitigate the problem of barren plateaus, or if it simply moves us from one region of barren plateaus caused by a certain dimension $d'$ to another region of barren plateaus caused by another dimension $d''$, where $d'' < d'$.

Therefore, in this article, besides identifying a new parameter influencing
barren plateaus, we also demonstrated that, although the use of VQAs with qudits may represent a promising path for solving various problems, their trainability and consequent applicability will be affected
by the dimension of the qudits. Moreover, the results obtained suggest that, although several mitigation methods have been proposed, their application in the context of VQAs using qudits still requires further investigation. This underscores the importance of continued research to address the challenges posed by barren plateaus in quantum machine learning algorithms utilizing qudits.

\vspace{0.3cm}

\begin{acknowledgments}
This work was supported by the Coordination for the Improvement of Higher Education Personnel (CAPES) under Grant No. 88887.829212/2023-00, the National Council for Scientific and Technological Development (CNPq) under Grants No. 309862/2021-3, No. 409673/2022-6, and No. 421792/2022-1, the National Institute for the Science and Technology of Quantum Information (INCT-IQ) under Grant No. 465469/2014-0, and by the S\~ao Paulo Research Foundation (FAPESP), Grant No. 2023/15739-3.
\end{acknowledgments}

\vspace{0.1cm}

\textbf{Data availability.} The code and data generated in this work are available at \url{https://github.com/lucasfriedrich97/BPQudit}.

\vspace{0.1cm}

\textbf{Contributions}  The project was conceive by L.F. and T.S.F.. The code was written by L.F., T.S.F.. J.M. was responsible for obtaining analytical results that assisted in optimizing part of the code. L.F. implemented the necessary numerical simulations to obtain the data for the results presented here. L.F., T.S.F., and J.M. were responsible for analyzing the obtained results. L.F., T.S.F., and J.M. jointly wrote the main manuscript.

\vspace{0.1cm}
\textbf{Competing interests.}
The authors declare to have no competing interests.


\begin{widetext}

\appendix
\section{Proof of Theorem 1}
\label{app}

In this appendix, we will present in details 
the proof of Theorem \ref{tr:1}.
To do this, we will start by reviewing some concepts mentioned briefly in the main text.


We will begin by introducing the concept of $t$-designs and discussing some lemmas that will be instrumental in deriving our theoretical results. Our goal here is not to offer an exhaustive review of this topic but rather to present the lemmas essential for our proofs. For a more comprehensive understanding, readers are encouraged to refer to Refs. \cite{Nakata2021, Chen2024}.

The formalism employed to elucidate the issue of barren plateaus relies on $t$-designs and the characteristics of the Haar measure. Let us consider a finite set $W_{y}$ of unitaries with a Hilbert space of dimension $d$, comprising $|Y|$ elements. If $P_{(t,t)}(W)$ represents an arbitrary polynomial of degree at most $t$ concerning the elements of the matrix $W$ and at most $t$ concerning those of $W^{\dagger}$, we define this finite set as a $t$-design \cite{2_design_SM}:
\begin{equation}
    \frac{1}{| Y |} \sum_{y =1}^{|Y|} P_{(t,t)}(W_{y}) = \int_{U(d)}d\mu(W)P_{(t,t)}(W),\label{eq:tdesign}
\end{equation}
where $U(d)$ signifies the unitary group of degree $d$. This outcome implies that the average of $P_{(t,t)}(W)$ over the $t$-design is essentially identical to integration across $U(d)$ with respect to the Haar distribution.

With this definition in mind, we can derive several lemmas that are pertinent to the analysis of the barren plateaus (BPs) issue, as discussed in the main text. Typically, in the literature, the dimension $d$ is straightforwardly defined as $d=2^{n}$, where $n$ signifies the number of qubits employed in the quantum circuit. However, when dealing with qudits, this dimension is expressed as $d=d'^{n}$, where $d'$ denotes the dimension of the qudits. Bearing this in consideration, we can establish the following lemmas.

\begin{lema}
\label{lema:1}
If $\{W_{y}\}_{y=1}^{|Y|} \subset U(d)$ forms a unitary $t$-design with $t \geqslant 1$, and let $A, B: H_{w} \rightarrow H_{w}$ be arbitrary linear operators, then we have:
\begin{equation}
\frac{1}{|Y|}\sum_{y =1}^{|Y|}\text{Tr}[W_{y}AW_{y}^{\dagger}B]  = \int d\mu(W)\text{Tr}[WAW^{\dagger}B] = \frac{\text{Tr}[A]\text{Tr}[B]}{d},
\end{equation}
where $d=d'^{n}$, and $n$ denotes the number of qudits.
\end{lema}

\begin{lema}\label{lema:2}
If $\{W_{y}\}_{y =1}^{|Y|} \subset U(d)$ forms a unitary $t$-design with $t \geqslant 2$, and let $A, B, C, D: H_{w} \rightarrow H_{w}$ be arbitrary linear operators, then we have:
\begin{equation}
\begin{split}
 \frac{1}{|Y|}\sum_{y =1}^{|Y|}Tr[W_{y}AW_{y}^{\dagger}B]Tr[W_{y}CW_{y}^{\dagger}D] & =  \int d\mu(W)Tr[WAW^{\dagger}B] Tr[WCW^{\dagger}D] \\
 & = \frac{1}{d^{2}-1}\big(Tr[A]Tr[C]Tr[BD]+ Tr[AC]Tr[B]Tr[D] \big)  \\
 & - \frac{1}{d(d^{2}-1)} \big( Tr[AC]Tr[BD]  + Tr[A]Tr[B]Tr[C]Tr[D] \big),
\end{split}
\end{equation}
where $d=d'^{n}$ with $n$ being the number of qudits.
\end{lema}

\begin{lema}\label{lema:3}
If $\{W_{y}\}_{y =1}^{|Y|} \subset U(d)$ forms a unitary $t$-design with $t \geqslant 2$, and let $A, B, C, D: H_{w} \rightarrow H_{w}$ be arbitrary linear operators, then we have:
\begin{equation}
    \begin{split}
         \frac{1}{|Y|}\sum_{y =1}^{|Y|}Tr[W_{y}AW_{y}^{\dagger}BW_{y}CW_{y}^{\dagger}D] & = \int d\mu(W)Tr[WAW^{\dagger}BWCW^{\dagger}D] \\
         & = \frac{1}{d^{2}-1}\bigg( Tr[A]Tr[C]Tr[BD] + Tr[AC]Tr[B]Tr[D] \bigg) \\ & - \frac{1}{d(d^{2}-1)} \bigg( Tr[AC]Tr[BD] + Tr[A]Tr[B]Tr[C]Tr[D] \bigg),
    \end{split}
\end{equation}
where $d=d'^{n}$ with $n$ being the number of qudits.
\end{lema}

As we have seen in the main text, the rotations gates for qudits can defined as:
\begin{equation}
    R_{jk}^{\alpha} = e^{-i\theta S_{\alpha}^{jk}/2}, \label{eq:si_rotationGate}
\end{equation}
where
\begin{align}
    S_x^{jk} &= |j\rangle\langle k| + |k\rangle \langle j|\\
    S_y^{jk} &= -i|j\rangle\langle k| + i|k\rangle \langle j|\\
    S_z^{j} &= S_z^{j0} = \sqrt{\frac{2}{j(j+1)}}\sum_{l=1}^{j+1} (-j)^{\delta(l,j+1)}|l\rangle \langle l|,
\end{align}
with the pair of variables $(j,k)$ in the matrices $S_x^{jk}$ and $S_y^{jk}$ obeying the relation $1 \leqslant j<k \leqslant d'$, and the variable $j$ in the matrix $S_z^{j}$ obeys the restriction $1 \leqslant j \leqslant d'-1$. Furthermore, it is useful to note that
\begin{equation}
    Tr[S_{\alpha}^{jk}] = 0 \label{eq:si_S}
\end{equation}
and
\begin{equation}
    Tr[(S_{\alpha}^{jk})^{2}] = 2. \label{eq:si_S2}
\end{equation}


As seen in the main text, the objective of a variational quantum algorithm (VQA) is to minimize a cost function $C$, typically defined as the average value of an observable $O$, such that
\begin{equation}
    C = Tr[OU \rho U^{\dagger}],
\end{equation}
with
\begin{equation}
    U = \prod_{l=1}^{L}U_{l}W_{l}
    \label{eq:si_U1}
\end{equation}
where 
\begin{equation}
    U_{l} = \bigotimes_{m=1}^{N}R_{jk}^{\alpha}(\theta_{ml}),
\end{equation}
with $R_{jk}^{\alpha}(\theta_{ml})$ defined in Eq. \eqref{eq:si_rotationGate}. Typically, we optimize the parameters $\pmb{\theta}$ of the parameterization $U$ using the gradient descent method. Our goal here is to obtain an expression for the partial derivative of the cost function with respect to any parameter. To do this, we begin by observing that the derivative of the cost function is given by:
\begin{equation}
    \partial_{k}C = \frac{\partial C}{\partial \theta_{pq}} = Tr \bigg[ O \bigg( [\partial_{k}U]\rho U^{\dagger} + U \rho [\partial_{k}U^{\dagger}] \bigg) \bigg].\label{eq:si_derC}
\end{equation}

To obtain $\partial_{k}U$, we initially rewrite Eq. \eqref{eq:si_U1} as:
\begin{equation}
    U = U_{L}U_{p}W_{p}U_{R},\label{eq:si_U2}
\end{equation}
with
\begin{equation}
    U_{L} = \prod_{l=1}^{p-1}U_{l}W_{l} \quad \text{ and } \quad U_{R} = \prod_{l=p}^{L}U_{l}W_{l}.
\end{equation}

As we are differentiating with respect to $\theta_{pq}$, from Eq. \eqref{eq:si_U2}, we will obtain:
\begin{equation}
    \partial_{k}U =  \frac{\partial U}{\partial \theta_{qp}} = U_{L}\bigg[\frac{\partial U_{p}}{\partial \theta_{qp}}\bigg]W_{p}U_{R},\label{eq:si_partialU}
\end{equation}
where we should remember that $W_{p}$ is a parameterization that does not depend on the parameters $\pmb{\theta}$. Now, as
\begin{equation}
    U_{p} = \bigotimes_{m=1}^{N}R_{jk}^{\alpha}(\theta_{mp}) = R_{jk}^{\alpha}(\theta_{1p}) \otimes \ldots \otimes R_{jk}^{\alpha}(\theta_{qp}) \otimes \ldots \otimes  R_{jk}^{\alpha}(\theta_{Np}),
\end{equation}
when differentiating with respect to $\theta_{pq}$, we will obtain
\begin{equation}
    \frac{\partial U_{p}}{\partial \theta_{qp}} = \bigotimes_{m=1}^{N}R_{jk}^{\alpha}(\theta_{mp}) = R_{jk}^{\alpha}(\theta_{1p}) \otimes \ldots \otimes \frac{\partial R_{jk}^{\alpha}(\theta_{qp})}{\partial \theta_{qp}} \otimes \ldots \otimes  R_{jk}^{\alpha}(\theta_{Np}).
\end{equation}

However, from Eq. \eqref{eq:si_rotationGate}, we obtain
\begin{equation}
    \frac{\partial R_{jk}^{\alpha}(\theta_{qp})}{\partial \theta_{qp}} = \frac{\partial e^{-i \theta S_{\alpha}^{jk}/2} }{\partial \theta_{qp}} = -\frac{i}{2} S_{\alpha}^{jk} R_{jk}^{\alpha}(\theta_{qp}),
\end{equation}
therefore 

\begin{equation}
    \begin{split}
    \frac{\partial U_{p}}{\partial \theta_{qp}} & = \bigotimes_{m=1}^{N}R_{jk}^{\alpha}(\theta_{mp}) = R_{jk}^{\alpha}(\theta_{1p}) \otimes \ldots \otimes \bigg( -\frac{i}{2} S_{\alpha}^{jk} R_{jk}^{\alpha}(\theta_{qp}) \bigg) \otimes \ldots \otimes  R_{jk}^{\alpha}(\theta_{Np}) \\
    & = \bigg[ I_{q} \otimes \bigg( -\frac{i}{2} S_{\alpha}^{jk} \bigg) \bigg]U_{p}.
    \end{split}
\end{equation}

Using this result in Eq. \eqref{eq:si_partialU}, we obtain

\begin{equation}
    \partial_{k}U = U_{L} \bigg( \bigg[ I_{q} \otimes \bigg( - \frac{i}{2} S_{\alpha}^{jk} \bigg) \bigg] U_{p} \bigg)W_{p}U_{R},
\end{equation}
or
\begin{equation}
    \partial_{k}U = U_{L} \bigg[ I_{q} \otimes \bigg( - \frac{i}{2} S_{\alpha}^{jk} \bigg) \bigg] U_{R}.
\end{equation}

Having obtained an expression for $\partial_{k}U$, now we can obtain an expression for $\partial_{k}C$. To do this, we simply need to substitute this result into Eq. \eqref{eq:si_derC}. Thus we have
\begin{equation}
    \begin{split}
        \partial_{k}C & = Tr \bigg[ O \bigg( U_{L}\bigg[ I_{q} \otimes \bigg( - \frac{i}{2} S_{\alpha}^{jk} \bigg) \bigg] U_{R} \rho U_{R}^{\dagger} U_{L}^{\dagger} +  U_{L}U_{R} \rho U_{R}^{\dagger} \bigg[ I_{q} \otimes \bigg( - \frac{i}{2} S_{\alpha}^{jk} \bigg)^{\dagger} \bigg] U_{L}^{\dagger} \bigg) \bigg] \\
        & = Tr \bigg[ O \bigg( -\frac{i}{2} U_{L} [I_{q} \otimes S_{\alpha}^{jk} ] U_{R} \rho U_{R}^{\dagger} U_{L}^{\dagger} + \frac{i}{2} U_{L}U_{R} \rho U_{R}^{\dagger} [I_{q} \otimes S_{\alpha}^{jk} ] U_{L}^{\dagger} \bigg) \bigg] \\
        & = \frac{i}{2} Tr \bigg[ O \bigg( U_{L}U_{R} \rho U_{R}^{\dagger} [I_{q} \otimes S_{\alpha}^{jk} ] U_{L}^{\dagger} -  U_{L} [I_{q} \otimes S_{\alpha}^{jk} ] U_{R} \rho U_{R}^{\dagger} U_{L}^{\dagger} \bigg) \bigg] \\
        & = \frac{i}{2} Tr \bigg[ U_{L}^{\dagger} O U_{L} \big[ U_{R}\rho U_{R}^{\dagger}, [I_{q} \otimes S_{\alpha}^{jk} ] \big] \bigg],
    \end{split}
\end{equation}
which is our expression for the partial derivative of the cost function $C$ with respect to any parameter $k$.


In this section, our objective is to present the proof of Theorem \ref{tr:1}. However, before doing so, we must demonstrate that 
\begin{equation}
    \langle \partial_{k}C \rangle = 0\  \forall k.\label{eq:si_med}
\end{equation}

This demonstration is relatively simple. To do this, we simply write
\begin{equation}
        \langle \partial_{k}C \rangle_{U_{R} U_{L}} = \frac{i}{2} \iint d \mu(U_{R}) d \mu(U_{L}) Tr \bigg[ U_{L}^{\dagger} O U_{L} \big[ U_{R}\rho U_{R}^{\dagger}, [I_{q} \otimes S_{\alpha}^{jk} ] \big] \bigg].\label{eq:si_medPartialC}
\end{equation}

Now, using Lemma \ref{lema:1}, we have that if $U_{L}$ forms a $1$-design, we obtain
\begin{equation}
        \langle \partial_{k}U \rangle_{U_{R} U_{L}} = \frac{i}{2} \int d \mu(U_{R}) \frac{1}{d} Tr[O] Tr \bigg[ \big[ U_{R}\rho U_{R}^{\dagger}, [I_{q} \otimes S_{\alpha}^{jk} ] \big] \bigg] = 0.
\end{equation}

If now, instead of solving the integral first with respect to $U_{L}$, we solve it with respect to $U_{R}$, we will obtain the same result. To do this, we just need to rewrite Eq. 
 \eqref{eq:si_medPartialC} as
\begin{equation}
        \langle \partial_{k}C \rangle_{U_{R} U_{L}} = \frac{i}{2} \iint d \mu(U_{L}) d \mu(U_{R}) Tr \bigg[ U_{R}\rho U_{R}^{\dagger} \big[ [I_{q} \otimes S_{\alpha}^{jk} ], U_{L}^{\dagger} O U_{L} \big] \bigg],
\end{equation}
where we used the cyclicity of the trace function. Now, if $U_{R}$ forms a $1-$design, we obtain, by using Lemma \ref{lema:1}, that
\begin{equation}
        \langle \partial_{k}C \rangle_{U_{R} U_{L}} = \frac{i}{2} \int d \mu(U_{L}) \frac{1}{d} Tr[\rho] Tr \bigg[  \big[ [I_{q} \otimes S_{\alpha}^{jk} ], U_{L}^{\dagger} O U_{L} \big] \bigg] = 0.
\end{equation}
This same result holds if $U_{L}$ and $U_{R}$ form a $1$-design simultaneously. Therefore, the result shown in Eq. \eqref{eq:si_med} is proven.

Now that we know that $ \langle \partial_{k}C \rangle = 0\ \forall k $, we need to obtain an expression for $Var[\partial_{k}C]$. However, since $Var[\partial_{k}C] = \langle (\partial_{k}C)^{2} \rangle - \langle \partial_{k}C \rangle^{2} $, we see that the variance will simply be given by
\begin{equation}
        \langle (\partial_{k}C)^{2} \rangle_{U_{R} U_{L}} = -\frac{1}{4} \int d \mu(U_{R}) d \mu(U_{L}) Tr \bigg[ U_{L}^{\dagger} O U_{L} \big[ U_{R}\rho U_{R}^{\dagger}, [I_{q} \otimes S_{\alpha}^{jk} ] \big] \bigg]^{2}.\label{eq:si_var1}
\end{equation}

Defining 
\begin{equation}
     \Gamma_{\alpha}^{jk} =  \big[ U_{R}\rho U_{R}^{\dagger}, [I_{q} \otimes S_{\alpha}^{jk} ] \big],\label{eq:si_var1.1} 
\end{equation}
we can rewrite Eq. \eqref{eq:si_var1} as
\begin{equation}
        \langle (\partial_{k}C)^{2} \rangle_{U_{R} U_{L}} = -\frac{1}{4} \int d \mu(U_{R}) d \mu(U_{L}) Tr[ U_{L}^{\dagger} O U_{L} \Gamma_{\alpha}^{jk} ] Tr[ U_{L}^{\dagger} O U_{L} \Gamma_{\alpha}^{jk} ].\label{eq:si_var2}
\end{equation}

Reorganizing the terms and using Lemma \ref{lema:2}, we obtain:
\begin{equation}
    \begin{split}
        \langle (\partial_{k}C)^{2} \rangle_{U_{R} U_{L}} = -\frac{1}{4} \int d \mu(U_{R}) \bigg[ & \frac{1}{d^{2}-1} \bigg( Tr[\Gamma_{\alpha}^{jk}]^{2}Tr[O]^{2} + Tr[(\Gamma_{\alpha}^{jk})^{2}]Tr[O^{2}] \bigg) \\ 
       & - \frac{1}{d(d^{2}-1)} \bigg( Tr[(\Gamma_{\alpha}^{jk})^{2}]Tr[O]^{2} + Tr[\Gamma_{\alpha}^{jk}]^{2}Tr[O^{2}] \bigg) \bigg].\label{eq:si_var3}
    \end{split}
\end{equation}
Since $Tr[\Gamma_{\alpha}^{jk}] = 0$, we obtain from Eq. \eqref{eq:si_var3}, after proper manipulations, that
\begin{equation}
    \langle (\partial_{k}C)^{2} \rangle_{U_{R} U_{L}} = \frac{1}{4} \bigg( \frac{ Tr[O]^{2} }{ d(d^{2}-1) } - \frac{Tr[O^{2}]}{ d^{2} -1 } \bigg) \int d \mu(U_{R}) Tr[(\Gamma_{\alpha}^{jk})^{2}].\label{eq:si_var4}
\end{equation}

Now we must solve the integral with respect to $U_{R}$. To do this, we must initially obtain an expression for $Tr[(\Gamma_{\alpha}^{jk})^{2}]$. Thus, from Eq. \eqref{eq:si_var1.1}, we have
\begin{equation}
    \begin{split}
        (\Gamma_{\alpha}^{jk})^{2} & = \bigg( \big[ U_{R}\rho U_{R}^{\dagger}, [I_{q} \otimes S_{\alpha}^{jk} ] \big] \bigg)^{2} \\
        & = \bigg( U_{R}\rho U_{R}^{\dagger} [I_{q} \otimes S_{\alpha}^{jk} ] - [I_{q} \otimes S_{\alpha}^{jk} ] U_{R}\rho U_{R}^{\dagger}  \bigg)^{2} \\
        & = U_{R}\rho U_{R}^{\dagger} A_{\alpha}^{jk} U_{R}\rho U_{R}^{\dagger} A_{\alpha}^{jk} - U_{R}\rho U_{R}^{\dagger} (A_{\alpha}^{jk})^{2} U_{R}\rho U_{R}^{\dagger} - A_{\alpha}^{jk} U_{R}\rho U_{R}^{\dagger} U_{R}\rho U_{R}^{\dagger} A_{\alpha}^{jk} + A_{\alpha}^{jk} U_{R}\rho U_{R}^{\dagger} A_{\alpha}^{jk} U_{R}\rho U_{R}^{\dagger},
    \end{split}
\end{equation}
where we define
\begin{equation}
    A_{\alpha}^{jk} = I_{q} \otimes S_{\alpha}^{jk}
    \label{eq:si_Ajk}
\end{equation}
in the last equality.

From this result, if we use the cyclicity of the trace operator, we obtain
\begin{equation}
    Tr[ (\Gamma_{\alpha}^{jk})^{2} ] = 2 Tr \big[ U_{R} \rho U_{R}^{\dagger} A_{\alpha}^{jk} U_{R} \rho U_{R}^{\dagger} A_{\alpha}^{jk} \big] - 2Tr \big[U_{R} \rho^{2} U_{R}^{\dagger} (A_{\alpha}^{jk})^{2} \big].
\end{equation}
So, using this result in Eq. \eqref{eq:si_var4}, we obtain
\begin{equation}
    \langle (\partial_{k}C)^{2} \rangle_{U_{R} U_{L}} = \frac{1}{2} \bigg( \frac{ Tr[O]^{2} }{ d(d^{2}-1) } - \frac{Tr[O^{2}]}{ d^{2} -1 } \bigg) \int d \mu(U_{R}) \bigg(  Tr \big[ U_{R} \rho U_{R}^{\dagger} A_{\alpha}^{jk} U_{R} \rho U_{R}^{\dagger} A_{\alpha}^{jk} \big] - Tr \big[U_{R} \rho^{2} U_{R}^{\dagger} (A_{\alpha}^{jk})^{2} \big] \bigg).\label{eq:si_var5}
\end{equation}

Now, using Lemmas \ref{lema:3} and \ref{lema:1} to solve the integral with respect to the first and second terms that appear in Eq. \eqref{eq:si_var5}, respectively, we obtain
\begin{equation}
    \begin{split}
        \langle (\partial_{k}C)^{2} \rangle_{U_{R} U_{L}} & = \frac{1}{2} \bigg( \frac{ Tr[O]^{2} }{ d(d^{2}-1) } - \frac{Tr[O^{2}]}{ d^{2} -1 } \bigg) \bigg[ \frac{1}{d^{2}-1} \bigg( Tr[\rho]^{2}Tr[(A_{\alpha}^{jk})^{2}] + Tr[\rho^{2}]Tr[A_{\alpha}^{jk}]^{2} \bigg) \\ & - \frac{1}{d(d^{2}-1)} \bigg( Tr[\rho^{2}]Tr[(A_{\alpha}^{jk})^{2}] + Tr[\rho]^{2}Tr[A_{\alpha}^{jk}]^{2} \bigg) - \frac{ Tr[\rho^{2}] Tr[(A_{\alpha}^{jk})^{2}] }{d} \bigg]
    \end{split}
\end{equation}
or
\begin{equation}
    \begin{split}
        \langle (\partial_{k}C)^{2} \rangle_{U_{R} U_{L}} & = \frac{1}{2} \bigg( \frac{ Tr[O]^{2} }{ d(d^{2}-1) } - \frac{Tr[O^{2}]}{ d^{2} -1 } \bigg) \bigg[ \frac{ Tr[A_{\alpha}^{jk}]^{2} }{ d^{2}-1 }\bigg( Tr[\rho^{2}] - \frac{Tr[\rho]^{2}}{d} \bigg) + \frac{ Tr[(A_{\alpha}^{jk})^{2}] }{d^{2}-1} \bigg( Tr[\rho]^{2}-dTr[\rho^{2}] \bigg) \bigg].\label{eq:si_var6}
    \end{split}
\end{equation}

At first, this result already informs us about how the variance behaves given $O$, $d'$ and $\rho$. However, since $\rho = | \psi \rangle \langle \psi |$ we have $Tr[\rho] = 1$. Furthermore, if we use $| \psi \rangle = \sum_{l} \beta_{l} | \beta_{l} \rangle $, with $ \sum_{l} \beta_{l} \beta_{l}^{*} = 1 $, we obtain
\begin{equation}
    \begin{split}
    Tr[\rho^{2}] & = Tr\bigg[ \bigg(\sum_{lm} \beta_{l} \beta_{m}^{*} | \beta_{l} \rangle \langle \beta_{m}| \bigg)^{2}  \bigg] =
    Tr\bigg[ \sum_{lmpq} \beta_{l} \beta_{m}^{*} \beta_{p} \beta_{q}^{*} | \beta_{l} \rangle \langle \beta_{m}|  \beta_{p} \rangle \langle \beta_{q}|   \bigg] \\
    & = Tr\bigg[ \sum_{lmpq} \beta_{l} \beta_{m}^{*} \beta_{p} \beta_{q}^{*} | \beta_{l} \rangle  \langle \beta_{q}| \delta_{m,p}   \bigg] =   \sum_{lmpq} \beta_{l} \beta_{m}^{*} \beta_{p} \beta_{q}^{*}  \delta_{m,p} \delta_{l,q}  \\
    & = \sum_{lm} \beta_{l} \beta_{m}^{*} \beta_{m} \beta_{l}^{*} =  \sum_{l} \beta_{l} \beta_{l}^{*} \bigg( \sum_{m} \beta_{m} \beta_{m}^{*} \bigg) = 1.
    \end{split}
\end{equation}

Therefore, we can rewrite Eq. \eqref{eq:si_var6} as
\begin{equation}
    \begin{split}
        \langle (\partial_{k}C)^{2} \rangle_{U_{R} U_{L}} & = \frac{1}{2} \bigg( \frac{ Tr[O]^{2} }{ d(d^{2}-1) } - \frac{Tr[O^{2}]}{ d^{2} -1 } \bigg) \bigg[ \frac{ Tr[A_{\alpha}^{jk}]^{2}
 }{ d(d+1) } -  \frac{ Tr[(A_{\alpha}^{jk})^{2}] }{d+1} \bigg].\label{eq:si_var7}
    \end{split}
\end{equation}
Finally, from Eq. \eqref{eq:si_Ajk}, we have
\begin{equation}
     Tr[A_{\alpha}^{jk}] = d'^{(n-1)} Tr[S_{\alpha}^{jk}]
\end{equation}
and
\begin{equation}
     Tr[(A_{\alpha}^{jk})^{2}] = d'^{(n-1)} Tr[(S_{\alpha}^{jk})^{2}].
\end{equation}

Therefore, from Eqs. \eqref{eq:si_S} and \eqref{eq:si_S2}, we obtain
\begin{equation}
    Tr[A_{\alpha}^{jk}] = 0 \quad \text{and} \quad Tr[(A_{\alpha}^{jk})^{2}] = 2 d'^{(n-1)}.
\end{equation}
Therefore, from Eq. \eqref{eq:si_var7} we obtain:
\begin{equation}
    \langle (\partial_{k}C)^{2} \rangle_{U_{R} U_{L}} = \frac{ d'^{(n-1)} }{d+1} \bigg( \frac{ Tr[O^{2}] }{ d^{2}-1 } - \frac{ Tr[O]^{2} }{ d(d^{2}-1) } \bigg).
\end{equation}
With this we complete the proof of Theorem \ref{tr:1}.

\section{Results in log-log scale}
\label{appB}

In this appendix, we show the same results from Section \ref{sec:result}, now represented using a log-log scale for the graphs. According to Definition \ref{def:2}, the BPs problem is amplified by the dimension of the qudits, and a polynomial decay of the variance with respect to the qudit dimension is observed. Furthermore, as demonstrated in Corollary \ref{cor:1}, when using the cost function defined in Eq. \eqref{eq:cost} with $O = |0\rangle \langle 0|$, the variance will decay polynomially with the qudit dimension. Therefore, the log-log scale allows us to empirically investigate whether this behavior is indeed observed.

In Figs. \ref{fig:model_1_log_log}, \ref{fig:model_2_log_log}, \ref{fig:model_3_log_log}, and \ref{fig:model_4_log_log}, in addition to replicating the results shown in the main text, we also plot the expected behavior of the variance according to Corollary \ref{cor:1}, here represented by the dashed line. This provides a visual reference for assessing whether the results obtained in the simulations align with the theoretical expectations. As observed, in general, the variance decreases as the qudit dimension increases, particularly as the depth of the parametrization used increases. In these cases, the observed behavior converges with the theoretical one, showing a polynomial decay of the variance as a function of the qudit dimension.

\begin{figure}[H]
    \centering
    \includegraphics[width=1\linewidth]{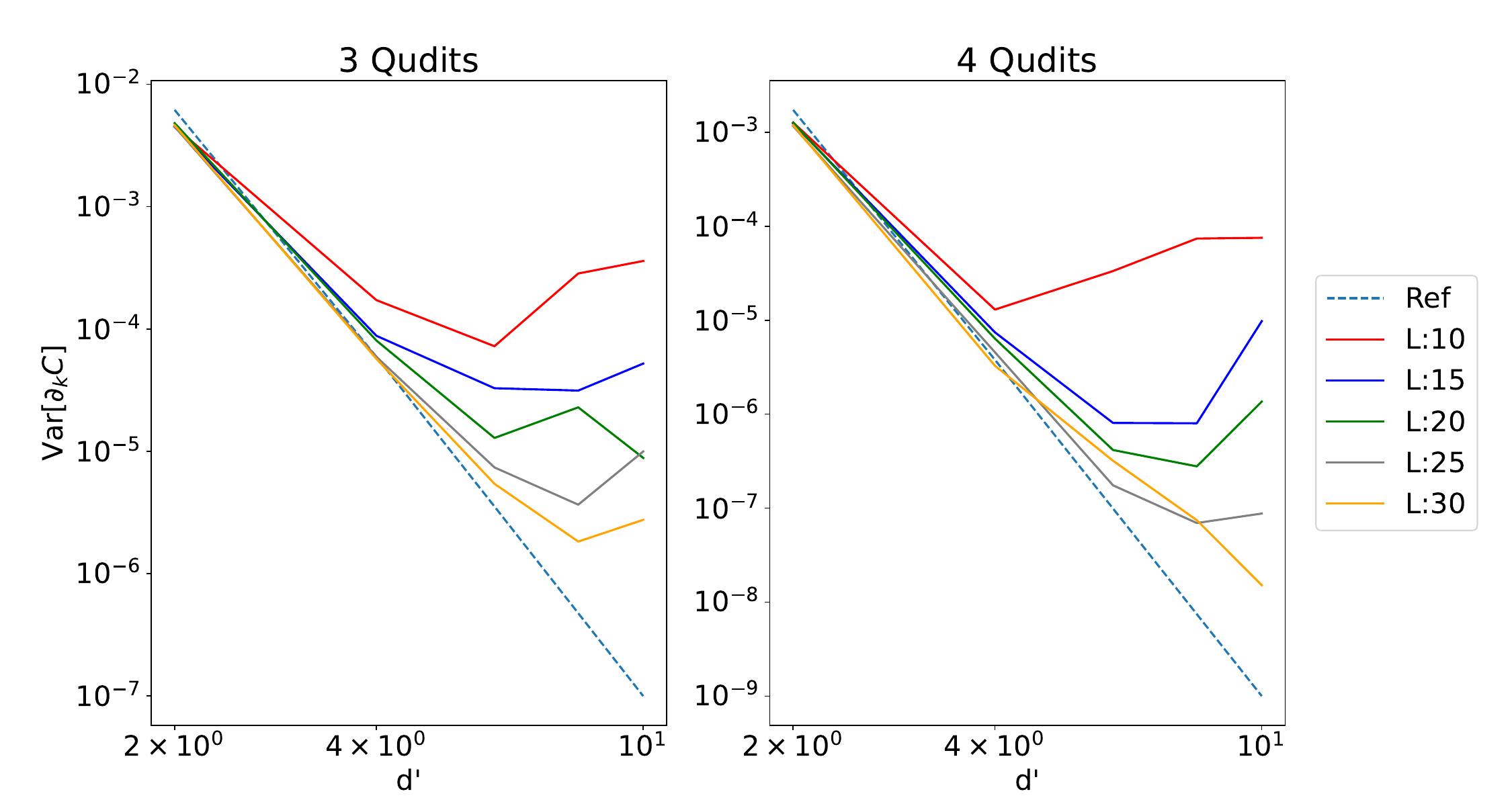}
    \caption{This figure presents the behavior of variance in relation to the dimension of the qudits using the parameterization presented in figure \ref{fig:param}A.  The dashed line shows the behavior of the expected variance according to Corollary \ref{cor:1}. }
    \label{fig:model_1_log_log}
\end{figure}

\begin{figure}[H]
    \centering
    \includegraphics[width=1\linewidth]{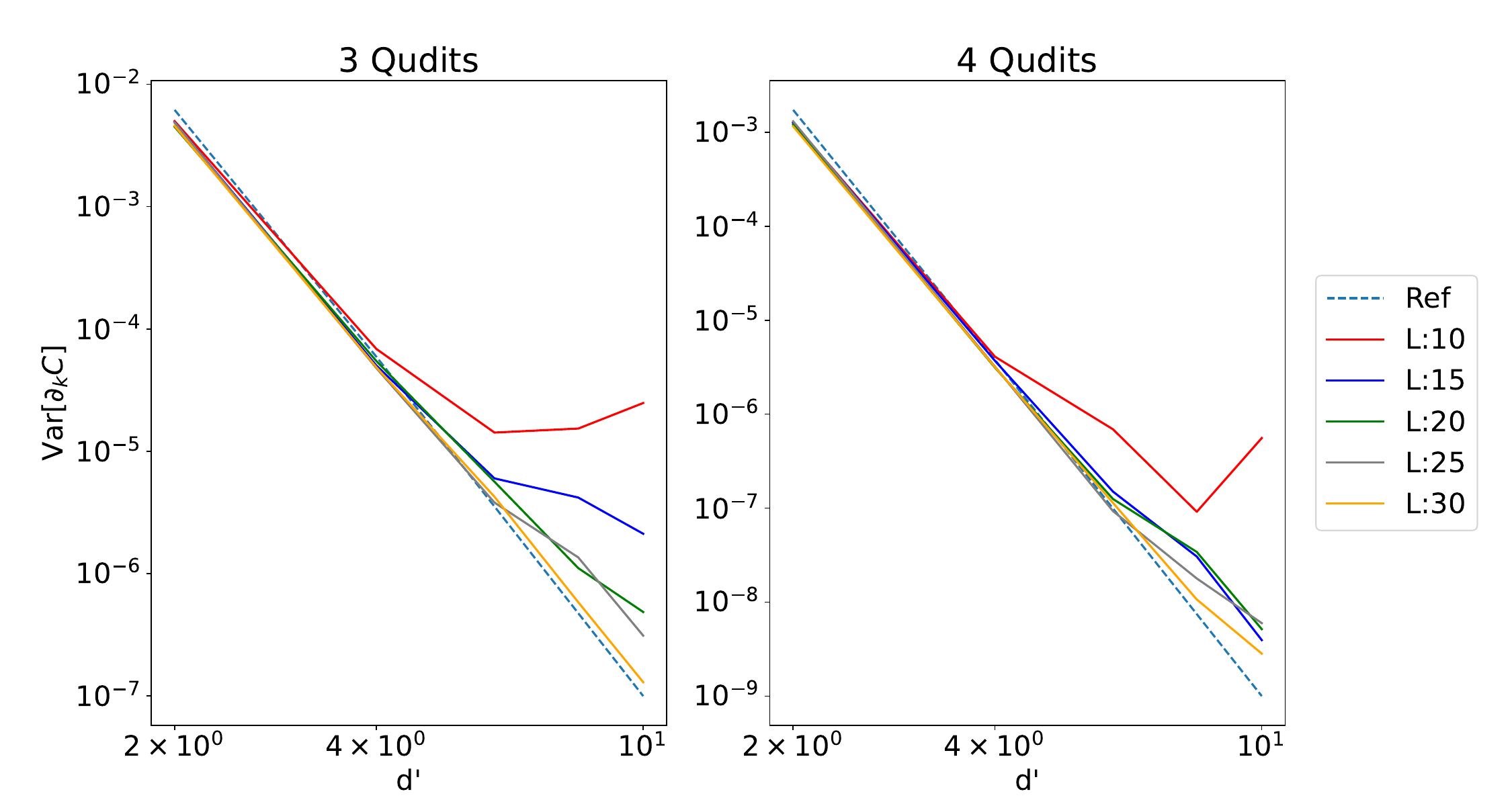}
    \caption{This plot illustrates the variance behavior as a function of the qudit dimension, based on the parameterization shown in figure \ref{fig:param}B. The dashed line represents the expected variance according to Corollary \ref{cor:1}.}
    \label{fig:model_2_log_log}
\end{figure}

\begin{figure}[H]
    \centering
    \includegraphics[width=1\linewidth]{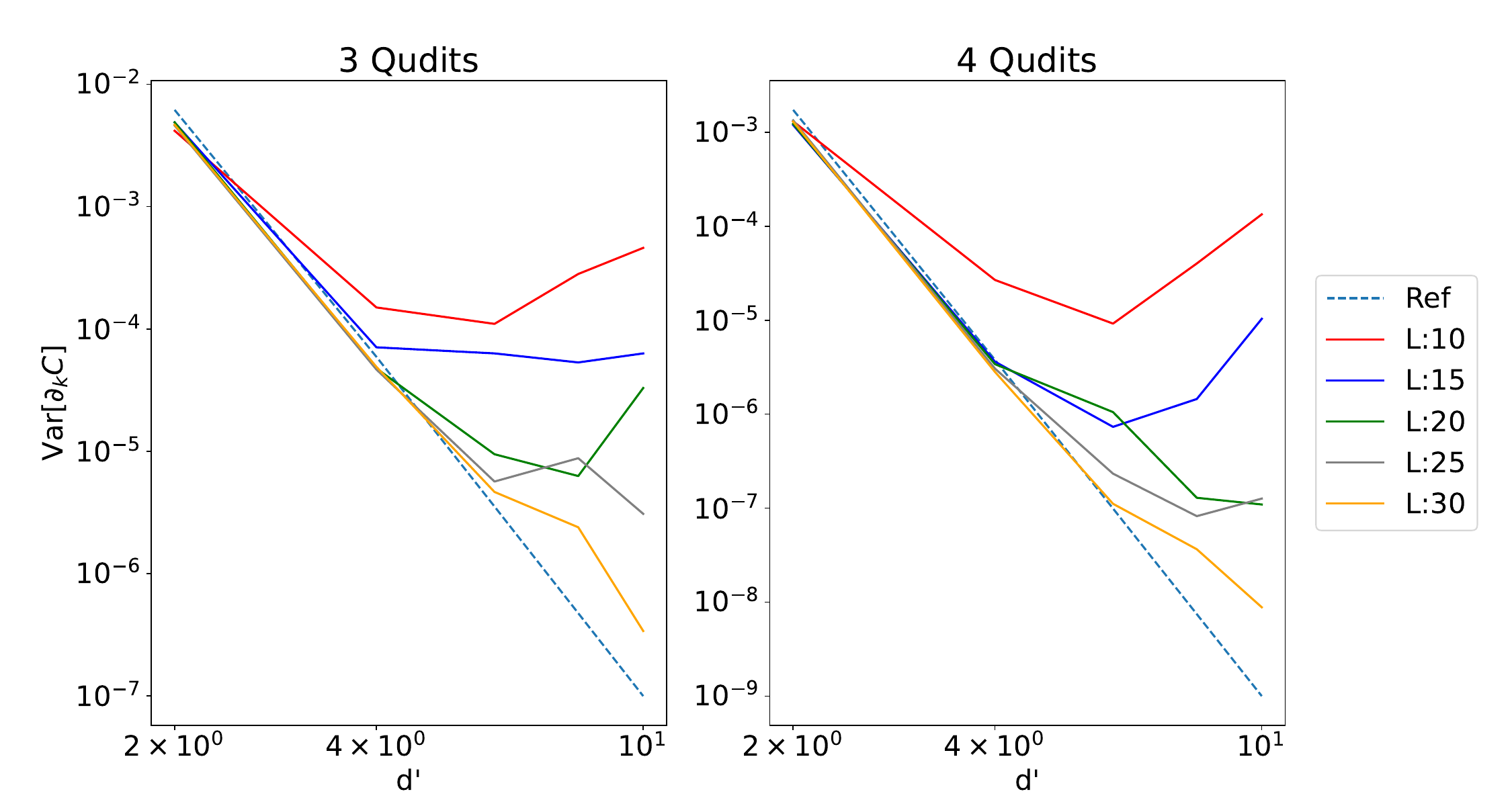}
    \caption{This figure depicts how the variance changes with the dimension of the qudits, considering the parameterization displayed in figure \ref{fig:param}C. The dashed line indicates the expected variance as stated in Corollary \ref{cor:1}.}
    \label{fig:model_3_log_log}
\end{figure}

\begin{figure}[H]
    \centering
    \includegraphics[width=1\linewidth]{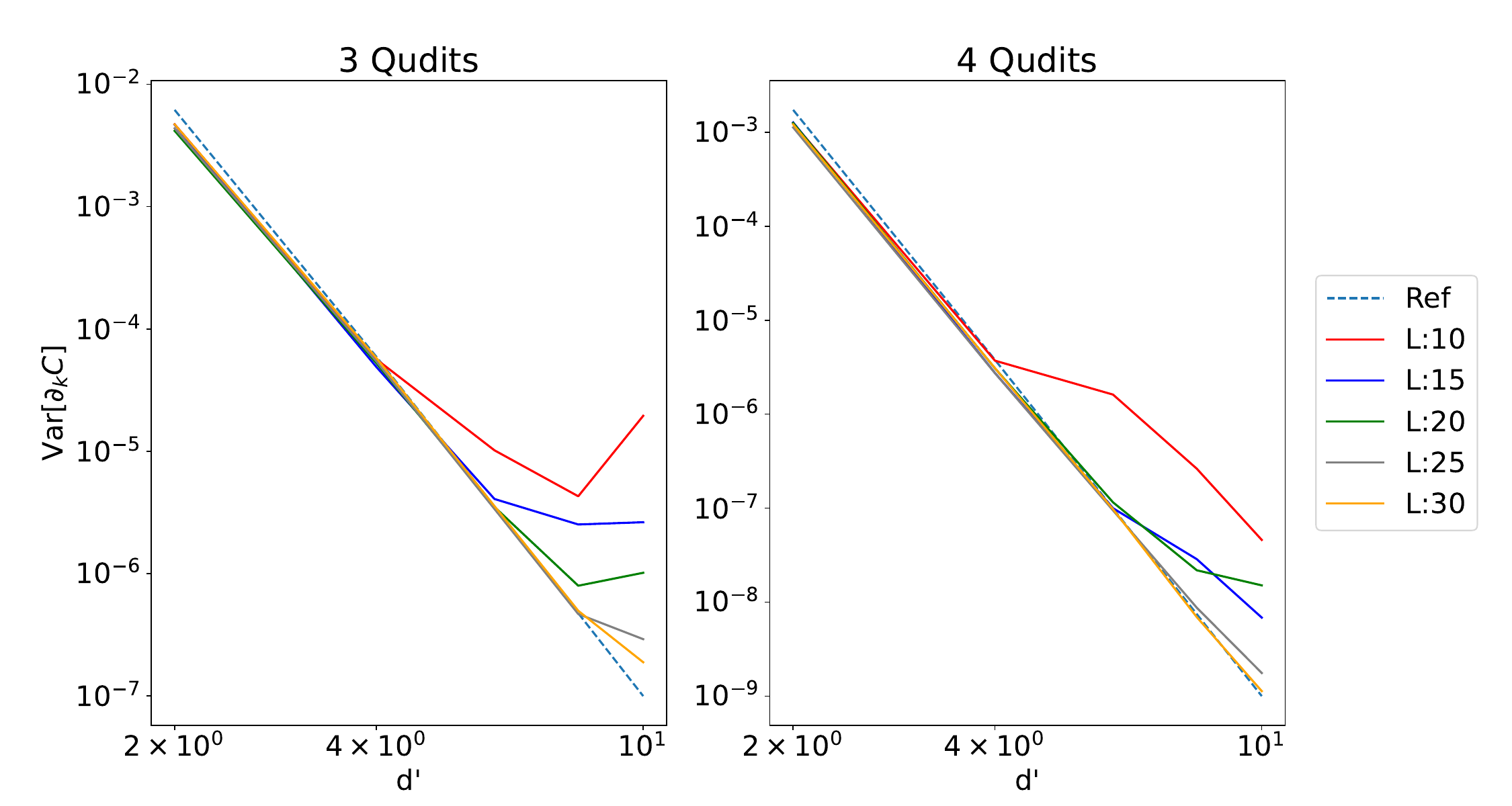}
    \caption{This chart shows the variance trend concerning the qudit dimension, following the parameterization outlined in figure \ref{fig:param}D. The dashed line highlights the expected variance according to Corollary \ref{cor:1}.}
    \label{fig:model_4_log_log}
\end{figure}
\end{widetext}

\clearpage

\end{document}